# Scalable Analysis for Arbitrary Photonic Integrated Waveguide Meshes

**Daniel Perez**[1*] **and Jose Capmany**[1,]

[1] *ITEAM Research Institute, Universitat Politècnica de València, Valencia, 46022, Spain.*
*\*Corresponding author: dperez@iteam.upv.es*

The advances in fabrication processes in different material platforms employed in integrated optics are opening the path towards the implementation of circuits with increasing degree of complexity. In addition to the more conventional Application Specific Photonic Circuit (ASPIC) paradigm the Programmable Multifunctional Nanophotonics (PMN) approach is a transversal concept inspired by similar approaches, which are already employed in other technology fields. For instance, in electronics Field Programmable Gate Array (FPGA) devices enable a much more flexible universal operation as compared to Application Specific Integrated Circuits (ASICs). In photonics, the PMN concept is enabled by two-dimensional (2D) waveguide meshes for which, the number of possible input/outputs ports quickly builds up and furthermore, internal signal flow paths make the computation of transfer functions an intractable problem. Here we report a scalable method based on mathematical induction that allows one to obtain the scattering matrix of any 2D integrated photonic waveguide mesh circuit composed of an arbitrary number of cells and which is easily programmable. To our knowledge this is the first report of the kind and our results open the path to unblock this important design bottleneck.

## 1. INTRODUCTION

Programmable Multifunctional Nanophotonics (PMN) aims at designing common integrated optical hardware configurations, which can implement a wide variety of functionalities by suitable programming [1-10] . Several authors [6,7,9,10] have reported theoretical work proposing different configurations and design principles based on the cascade of either beamsplitters [7,9, 10] or integrated Mach Zehnder Interferometers [6] (MZIs). A more versatile architecture can be obtained by following similar principles as those of the Field Programmable Gate Arrays (FPGAs) in electronics [1-5]. The core concept is to break down complex circuits in a large network of identical unit cells implemented by means of a two-dimensional (2D) waveguide mesh or lattice. Different functionalities are then achieved by selecting the adequate path through the mesh. 2D integrated waveguide meshes formed by replicating a square [2], hexagonal [1, 3,5] or triangular [1] unit cells provide the required regular and periodic geometries, where each side of the basic cell is implemented by two waveguides coupled by an independent (power splitting and phase) tunable basic unit (TBU). Several simple configurations with a reduced number of cells (i.e up to 7) have been recently reported, providing solid proofs of concept [3] and demonstrating the capability of implementing both traditional signal processing architectures as well as arbitrary linear matrix transformations which are at the heart of most applications targeted for photonic chips. For instance, in quantum information, NxN unitary transformations support the implementation of simple and complex logic gates [11-17], the emulation of boson sampling [18-20] circuits and quantum lab on a chip [21], to cite a few applications. Waveguide meshes open the path for reconfigurable large-scale integrated quantum information systems with a potential to supersede current approaches based on static configurations [22]. In computer processor interconnections, reconfigurable broadband inter-processor and computer interconnections are fundamental in high-performance computing and data centers [23]. Photonic linear transformations provide a clean, interference-free and high-speed option for core processor resource management [24]. In optical signal processing, linear transformations that can be supported by PMN processors based on 2D waveguide meshes include several operations that are central to optical signal processing as, for example: the optical FFT [25], Hilbert transformation [26], Integrators and differentiators [27,28]. In Neurophotonics, unitary (NxM) and non-unitary (NxM) matrix transformations are fundamental building blocks preceding nonlinear threshold operations in neural networks, spike and reservoir computing [29,30]. The availability of PMN processors opens an interesting and exciting research avenue in this emerging field. In biophotonic sensing, PMNs support of simple and Multiple Input/Multiple Output (MIMO) interferometric structures for lab-on-a-chip enabling the future implementation of multi-parameter integrated photonic sensing [31,32]. Finally, but not least important, in advanced physics, waveguide mesh PMN provides a programmable 2D platform to implement different topological systems such as multi-ring cavity structures to support research in synthetic dimensions [33] and devices based on topological insulator principles [34,35].

The extension of 2D waveguide meshes to account for an increased number of cells and therefore to implement more complex structures including a higher number of TBUs (>80) is desirable, as this will dramatically expand the number of functionalities that can be implemented with a given hardware configuration. Several physical and design limitations have to be overcome and one of the most important is the availability of a scalable method of analysis [2,3,5]. A correct spectral characterization calls for a scattering matrix [36-38] method, but the main difficulty resides in the fact that the input/output port count and the internal interconnections in the 2D structure suffers very steep increase with the number of cells in the chip, making in appearance this problem analytically intractable.

Mathematical induction (MI), is a technique that can be employed to prove some particular rule or pattern, usually infinite or arbitrarily large [39]. It is based on two steps, the base step where a simple case is established and an induction step, which involves showing that an arbitrary large example follows logically from a slightly smaller one. In mathematical terms the principle of induction states that for a fixed integer b and for each integer $n≥b$, let S(n) be a statement involving n. If (i) $S(b)$ is true and (ii) for any integer $k≥b$, $S(k)→S(k+1)$ then for all $n≥b$, the statement $S(n)$ is true. This apparently simple principle conceals in fact a very strong proof technique that finds applications in a myriad of fields including [39] probability, geometry, game theory, graph theory, systems complexity, and artificial intelligence. In particular MI is very attractive for probing very general and powerful results about large count and infinite structures.

Here we report a scalable method based on mathematical induction that allows one to obtain the analytic scattering matrix of any 2D integrated photonic waveguide mesh circuit composed of an arbitrary number of cells and which is easily programmable. To our knowledge this is the first report of the kind and our results open the path to unblock this important design bottleneck. The method not only provides all the desired input/output transfer functions, but also allows to design the unused regions of the waveguide mesh so they can be employed to manage undesired contributions from reflected and crosstalk signals and thus optimize the chip performance and furthermore, it allows to study all the input/output responses as the internal parameters of the TBUs are changed opening the path for error evaluation via Monte Carlo simulations and the incorporation of machine learning algorithms for circuit self-correction. Although the procedure proposed here is developed for a hexagonal waveguide mesh, it can be applied to any uniform 2D mesh topology.

## 2. PROBLEM FORMULATION AND INDUCTIVE METHOD

In photonics, the PMN concept is enabled by two-dimensional waveguide meshes formed by replicating square, hexagonal [1,3,5] or triangular [1] unit cells. Each side of the unit cell is implemented by two waveguides coupled by an independent (power splitting and phase) tunable basic unit (TBU). This element can be implemented by means of tunable 3-dB MZIs or by a dual-drive directional coupler and described by a *2x2 transmission matrix* $H_{TBU}$. In the case of hexagonal waveguide meshes the basic building block or trilattice is formed by three TBUs (A, B and C) connected in a Y configuration as shown in Figure 1.a. The trilattice is described by a *6x6 scattering matrix* computed from the three scattering matrices HTBU describing its internal TBUs. To aid in the graphical illustration of the method we will employ a triangle symbol to represent the trilattice, where each port has, in principle, internal connections to the four opposite ports (i.e port 1 to ports 3,4,5,6, etc). The trilattice can be replicated and distributed *N*-times to generate any desired hexagonal mesh arrangement of any size. For example, Figures 1b and 1c show the process leading to the construction of a single a hexagonal cell composed of three tri-lattices (we employ the notation $A_i$, $B_i$, $C_i$ to identify the TBUs that compose trilattice i). Even for the simplest structure representing the unit cell one has already 12 input/output ports and 6 intermediate auxiliary nodes required for the computation of the *12x12* (i.e 144 element) transfer matrix. With increasing number of cells the above figures show a drastic increase. For instance, the 4-cell structure shown in figure 1.d, which is still a low complexity structure features 20 input/output ports, 38 internal nodes and a *20x20* (i.e 400 element) scattering matrix. Figure 1.e provides the exact input/output port count and the internal nodes as a function of the number of hexagonal cells. It clearly shows that the analytic derivation of scattering matrices for 2D meshes becomes apparently intractable even for a very low cell count. Moreover, numerical methods to analyze circuit responses such as finite-difference time domain FDTD and eigenmode solvers do not scale well as the number of components in the photonic circuit increase.

We propose here a method for the analytic determination of the full scattering matrix of waveguide meshes composed of an arbitrary number of hexagonal cells. The method uses mathematical induction and is based on increasing 2D hexagonal waveguide meshes formed by *n-1* trilattices by adding an extra trilattice unit. Formally the method is stated in the following way. A 2D structure formed by one trilattice is described by a unitary scattering matrix H(1) with known coefficients. Then, if a 2D structure formed by *n-1≥1* trilattices is described by a unitary scattering matrix H(n-1) with known coefficients, the structure composed of n trilattices obtained from appending an additional trilattice H(1) to the former is described by a unitary scattering matrix H(n) with known coefficients.

This method allows the sequential derivation of the scattering matrix of a n-th order arbitrary hexagonal waveguide mesh using the scattering matrix of the previous lower order mesh H(n-1) and that of the newly added trilattice H(1). Its final computation will depend on how the additional trilattice is connected to previous lower order mesh. Four different interconnection scenarios can be identified, as shown in figure 2.a to 2.d depending on the number of ports that are interconnected and the number of new complete hexagonal cells that appear after incorporation of the new tri-lattice:

*Scenario 0:* This is the simplest case and the starting point in the generation of a new mesh design. Here, only one out of the 6 ports that define the tri-lattice is connected to the previous mesh ports. The addition of the new tri-lattice increases the number of mesh ports by 4, increasing the number of rows and columns in the scattering matrix, correspondingly.

*Scenario 1:* Here, the addition of the new tri-lattice increases the number of mesh ports by two but the number of complete hexagonal cells does not increase.

*Scenario 2:* Here, the addition the new tri-lattice increases the number of ports by two and the number of complete cells by one.

*Scenario 3:* In this case, the addition the new tri-lattice does not increase the number of ports, since it connects three ports to the previous mesh and the number of complete cells is increased by one.

Figure 2.e illustrates the most general signal flow diagram that needs to be taken into account for deriving H(n) as a function of H(n-1) and H(1). For each scenario this diagram can be simplified as shown in the supplementary material. Nodes *s*, and *r* shown in pink in the left hand-side represent any pair of input and output ports respectively (the allowed ranges of variation for *s* and *r* are also shown depending on the scenario, where P is the input/output port count of H(n-1) before the connection of the additional trilattice). Nodes *x, y* and *z* identify the input/output ports of H(n-1) that are employed to connect this mesh to the newly added trilattice (the allowed values for *x, y* and *z* are also shown depending on the scenario). Connections *N, M, X, Y, F, D E', F', Q, R, C', D', A', B', S, U, I, J, B, F*, $h_{yy}$, $h_{zz}$, $h_{xx}$ represent signal flow paths with transfer functions given by coefficients of the scattering matrix H(n-1). Connections *K, L, O, P, A, H, C, E, T, G, V, W* represent the additional signal flow paths that result from the additional trilattice. The transfer functions (additional matrix coefficients) for these connections must be computed in order to obtain the overall scattering matrix $H(n)$ (the derivation of these matrix elements is provided in the supplementary material for the four different scenarios).

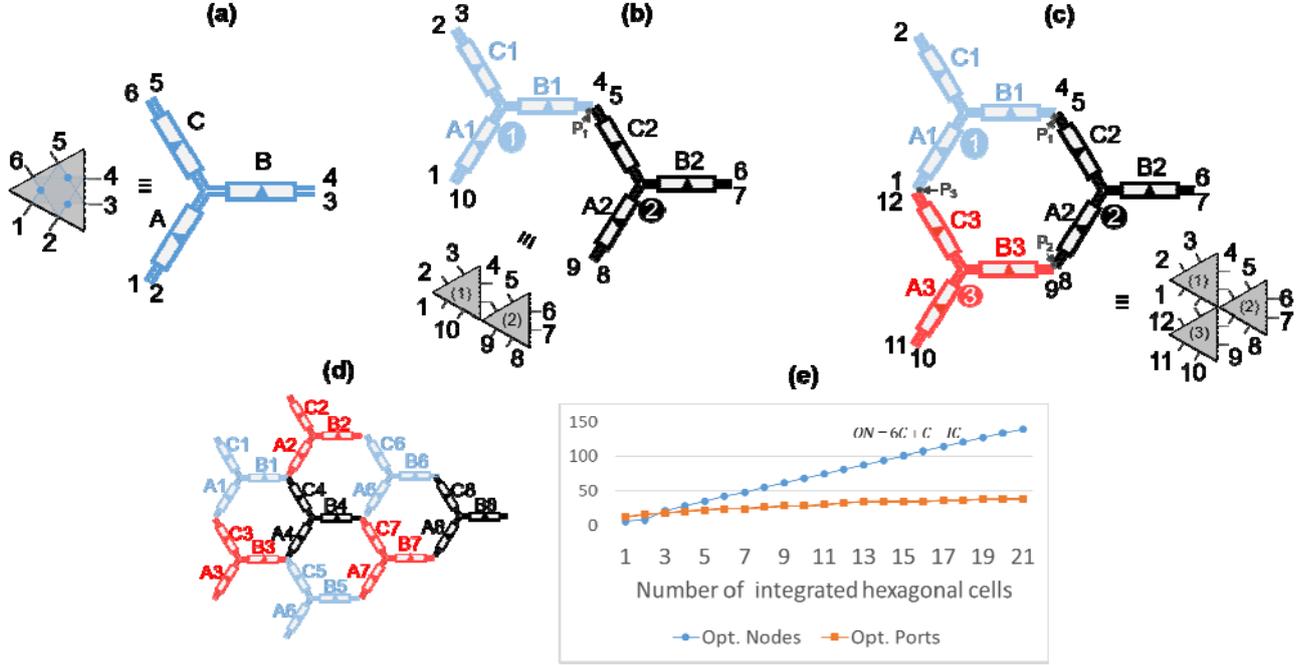

**Fig. 1 |** Trilattice building block for 2D hexagonal waveguide meshes and increased complexity in the required number of input/output and internal nodes- (a), trilattice composed of three TBUs and associated symbol, (b), two trilatteces interconnected by the optical node P1, (c) Three trilattices creating a closed hexagonal cell, (d), eight trilattices interconnected to obtain a four-cell count waveguide mesh. (e), number of optical nodes (ON) and optical ports versus number of closed cells (C) in a waveguide mesh photonic integrated circuit IC in the equation stands for the number of closed cells surrounded by closed cells.

## 3. RESULTS

To illustrate the method of analysis we have chosen a waveguide mesh composed by 18 hexagonal cells built upon assembling 27 trilattices as shown in figure 3.a. This structure, which has around twice the number of cells than those corresponding to the current state of the art, features 40 input/output ports and 122 intermediate ports. Input/output ports are numbered clockwise. To estimate the order of magnitude involved and the complexity in characterizing this mesh, each programming results in a *40x40* matrix with up to 1600 elements for each operation wavelength. In other words, *1600 x $W_p$* potential complex-valued transfer functions when the optical spectrum is swept over $W_p$ wavelength points.

### A. Single wavelength Analysis

We first show the application of the method when a single wavelength operation is considered. We programmed the mesh to implement two multiport rectangular interferometers simultaneously. A *3x3* interferometer is shown inside a blue box, while a 4x4 interferometer is shown in a yellow box. Figure 3.b shows the circuit layouts for the 3x3 and 4x4 interferometers including the input and output ports in the mesh structure shown in red ink. The coupling factors and phases of the TBUs emulating the interferometers are programmed to implement several transformations (see the Supplementary Material 6 for the coefficient tables of the implemented circuits). In the first case the mesh implements a unitary 3x3 discrete Fourier transform, and a non-unitary 2x4 hybrid commonly employed in coherent receivers.

$$H_{DFT3} = \frac{1}{\sqrt{3}} \begin{pmatrix} 1 & 1 & 1 \\ 1 & e^{i2\pi/3} & e^{i2\pi/3} \\ 1 & e^{-i2\pi/3} & e^{i2\pi/3} \end{pmatrix}$$

$$H_{Hybrid} = \frac{1}{2} \begin{pmatrix} 1 & 1 \\ 1 & -1 \\ 1 & i \\ 1 & -i \end{pmatrix}$$

(1)

The results obtained using the analytic method for a single wavelength are shown in figure 3.c. both modulus and phase of the 1600 matrix coefficients are displayed in a bi-dimensional map relating the input and output ports. In each map the input/output connections below the diagonal show the left-to-right direction of propagation while the input/output connections above the diagonal show the right-to-left direction of propagation. The matrix elements limited by the broken-dotted rectangles correspond to the desired transformations. The method retrieves the transformation matrices given by equation (1). In addition, the non-desired paths established between input and output ports that result from the programming of the waveguide mesh TBUs are also obtained as non-zero matrix coefficients.

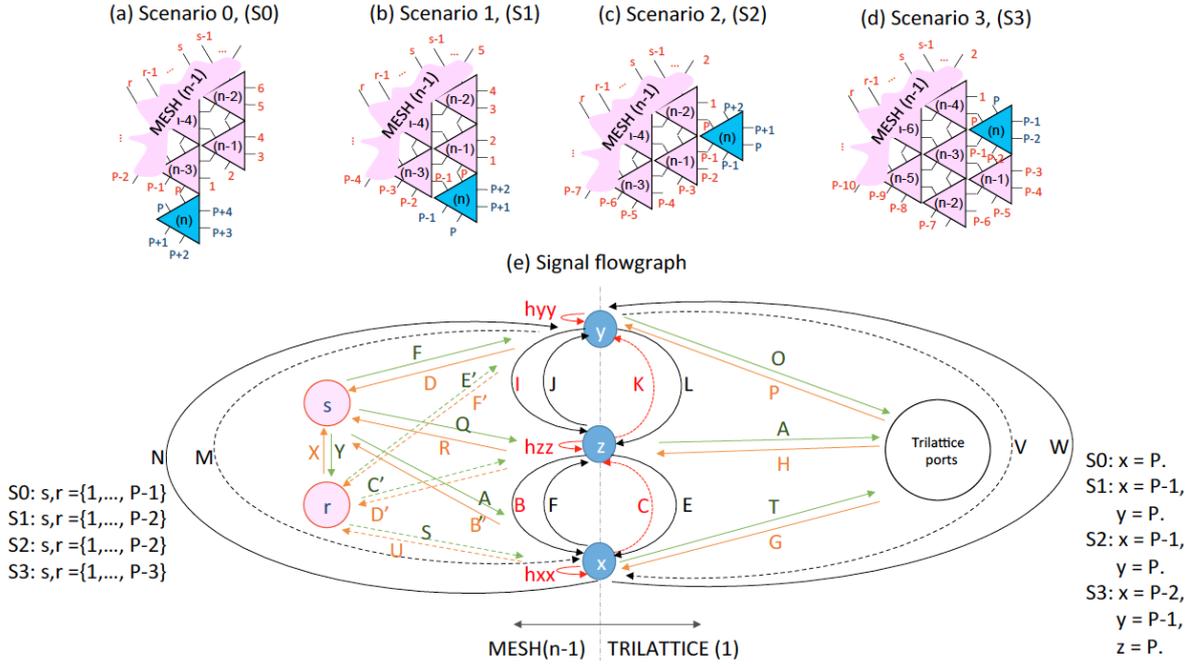

**Fig. 2** | Inductive method description for obtaining the scattering matrix $H(n)$ of an hexagonal 2D waveguide mesh composed of $n$ basic trilattice units by addition of one trilattice unit $H(1)$ to an hexagonal 2D waveguide mesh composed of $n-1$ basic trilattice units $H(n-1)$ and general signal flowgraph for its implementation. a, Interconnection scenario 0. b, Interconnection scenario 1. c, Interconnection scenario 2. d, Interconnection scenario 3. e, General signal flow diagram that needs to be taken into account for deriving $H(n)$ as a function of $H(n-1)$ and $H(1)$. Connections $N, M, X, Y, F, D\ E', F', Q, R, C', D', A', B', S, U, I, J, B, F, h_{yy}, h_{zz}, h_{xx}$ represent signal flow paths with transfer functions given by coefficients of the scattering matrix $H(n-1)$. Connections $K, L, O, P, A, H, C, E, T, G, V, W$ represent the additional signal flow paths that result from the additional trilattice.

The latter have no impact over the circuit operation provided that no input signal is fed to these undesired ports. However, some of these can still be employed and input signals will have no impact over the programmed circuits. This information is provided in the 2D matrix maps. For instance, 3.c. shows that no input signal can be allowed in ports 12 and 13 for the correct operation of the hybrid while input signals from ports 3, 5, 7, 9 etc. will have no impact.

In the second case the mesh is programmed to implement a 3x3 beamsplitter and a 4x4 Hadamard transformation.

$$H_{Tritter} = \frac{1}{\sqrt{3}} \begin{pmatrix} 1 & 1 & 1 \\ 1 & e^{j2\pi/3} & e^{j4\pi/3} \\ 1 & e^{j4\pi/3} & e^{j8\pi/3} \end{pmatrix}$$

$$H_{Had} = \frac{1}{\sqrt{2^2}} \begin{pmatrix} 1 & 1 & 1 & 1 \\ 1 & -1 & 1 & -1 \\ 1 & 1 & -1 & -1 \\ 1 & -1 & -1 & 1 \end{pmatrix} \tag{2}$$

The results rendered by the analytic method are displayed in figure 3.d. again showing an excellent degree of matching with the desired transformations given by (2). The use of non-ideal components has been considered in the Supplementary Material Note 3, together with more examples of linear matrix transformations.

**B. Full Spectral Analysis**

The more powerful and versatile characteristics of the analytic method are however unleashed when using it for spectral characterization. Here, the wavelength (or frequency) dimension is added and truly spectral transfer functions are immediately provided in a few seconds. Figure 4.a shows as an example, the layout of the programmed waveguide mesh to implement a Side Coupled Integrated Spaced Sequence of Resonators (SCISSOR) filter composed of 5 cascaded ring resonators with 6-BUL cavity length (see Supplementary Material 6 for the programming table of the phase shifters). Figure 4.b. shows the emulated circuit layout, where the input port (16) and the output port (34) are marked in red ink. Within each ring cavity one TBU in bar state (B4, B6, B13, B15 and B22) is employed as phase shifter to provide additional resonance displacements ($\phi_1, \phi_2, \phi_3, \phi_4, \phi_5$) if required with respect to that set by the cavity free spectral range. The ring couplers are implemented by TBUs set in tunable coupler mode (B3, B7, B12, B16, B21). For practical reasons all of them are set to provide the same coupling constant $K$. The input signal vector is $I=(i_1,i_2,...i_{40})$ where $i_k=0$ unless $k=16$ and $i_{16}=1$. The upper trace of Figure 4.b plots the moduli of the 40 transfer functions obtained after multiplying the 40x40 matrix of spectral transfer functions (each transfer function is computed for 1001 different wavelengths) by the input vector I for the case where $K=0.2$ and $\phi_1=\phi_2=\phi_3=\phi_4=\phi_5=0$. Note that the wavelength axis is normalised to the length of a single TBU. As expected, only one transfer function (corresponding to the matrix coefficient $h_{34,16}$) is relevant, while the other 39 represent noisy contributions due to the undesired signal leakage from the input port 16 to the rest of the output ports. Note that an optical crosstalk of -20 dB has been assumed for this example. This information is however very useful as it will be shown later when optimizing the mesh performance to decrease the impact of crosstalk. The phase response of $h_{34,16}$ is shown in detail in the intermediate trace of Figure 4.b., while the lower trace illustrates the spectral response $h_{34,16}$ for two different cases where the SCISSOR parameters are changed. In the first (red trace), $K=0.2$ in all the rings and the ring resonances are slightly detuned $\phi_1=-0.12, \phi_2=-0.06, \phi_3=0, \phi_4=0.06, \phi_5=0.12$, to reduce the filter bandpass and main to secondary sidelobes. In the second (yellow trace), the coupling constants are apodized $K_1=0.39, K_2=0.47,$

$K_3=0.55$, $K_4=0.63$, $K_5=0.71$, and the ring resonances strongly detuned $\phi_1=-0.4$, $\phi_2=-0.2$, $\phi_3=0$, $\phi_4=0.2$, $\phi_5=0.4$. The obtained results using the inductive method are in exact coincidence with those resulting from typical matrix multiplication of ring cavities. Note that while having a single input signal fed to the mesh activates only 40 input-output responses, the method directly calculates 1600 potential transfer functions, each of them with 1000 spectral points in few seconds. (See Supplementary Material 4 for the programming of a wider set of examples)

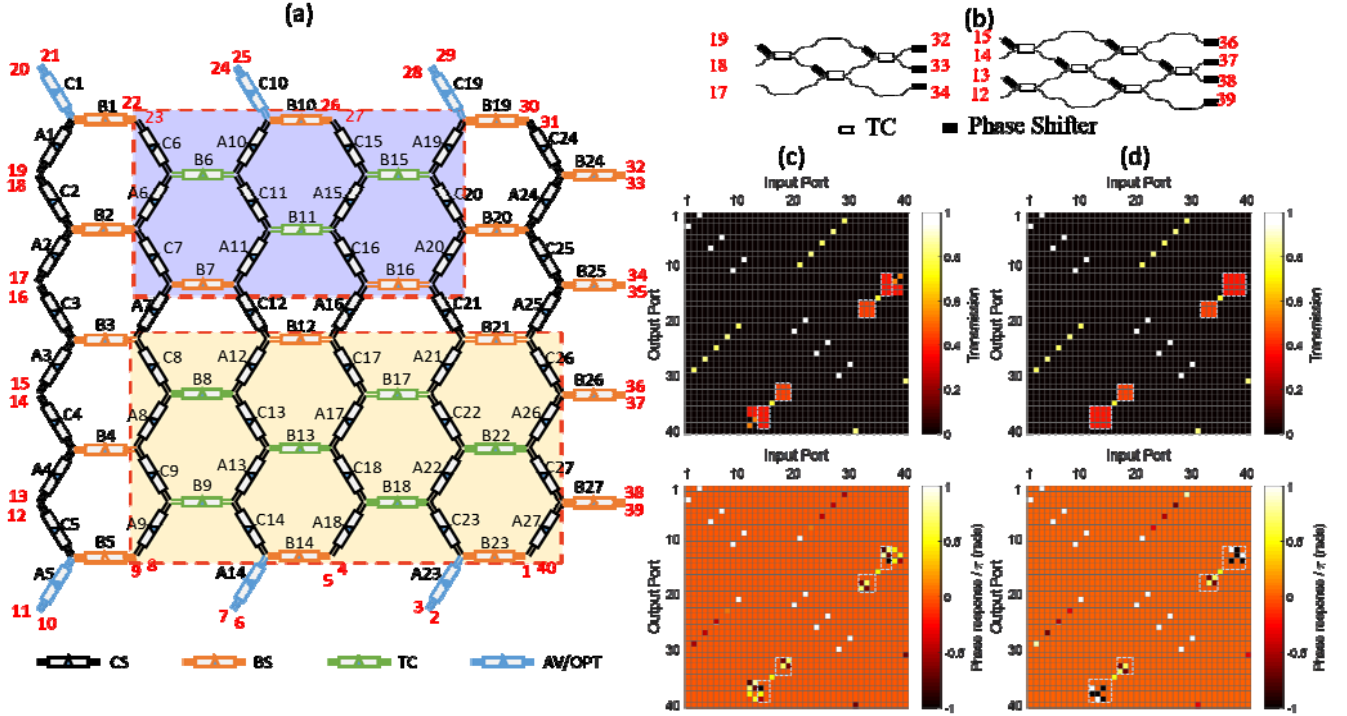

**Fig. 3** | Scalable Analysis Method application to single wavelength operation of waveguide mesh configuration for universal linear interferometers. (a) Mesh architecture and configuration for simultaneously implementing *3x3* and *4X4* linear transformations. (b) Equivalent circuit layouts with indication of the input and output ports in red ink. (c) Moduli and phases of all the *40x40* matrix coefficients when the *3x3* and *4x4* transformations are programmed to implement a DFT and a *2x4* Optical Hybrid respectively. (d) Moduli and phases of all the *40x40* matrix coefficients when the *3x3* and *4x4* transformations are programmed to implement a Three-way beamsplitter and a *4x4* Hadamard matrix respectively.

### C. Multiparameter Error Analysis

The most important sources of impairment in the operation of waveguide mesh circuits derive from the fact that either imperfect components are obtained as a result of the fabrication process or that the setting values for the structure phase shifters depart from the ideal values required by design [40-42]. In either case these result in deviations from the targeted circuit performance. Typical errors in component fabrication are connected with departure from the 50/50 power splitting ratio of 3 dB couplers employed to implement MZI based TBUs and the imprecise settings of the waveguide mesh phase shifters. To evaluate the impact that this effect has on the circuit performance one has to resort to Monte Carlo analysis, where the operation of each TBU can be modeled by two Gaussian random variables centered at its ideal setting and featuring a standard deviation $\sigma_K$ and $\sigma_\phi$ accounting for random fluctuations around the mean of the coupling coefficient ($K$) and phase term ($\phi$), respectively. (See Supplementary Material 5 for the TBU modeling details.) This process is quite time consuming and has been applied, to our knowledge, only to feed-forward waveguide mesh circuits with a certain degree of complexity [41,42]. Our analytic method speeds-up this analysis allowing for 1000 realizations in a few seconds. Furthermore, it is applicable both to feed-forward and feed-backward circuits as both can be emulated by the waveguide mesh. As an example, figure 5 shows the results of a Monte Carlo analysis for a feed-forward/feed-backward circuit. The mesh is programmed to implement a double ring loaded MZI filter that is employed to implement maximally-flat passband Butterworth and Bessel type filter [43, 44]. Figure 5.a. depicts the waveguide mesh programming to implement the circuit layout shown in Figure 5.b. Phase shifter coefficients for ideal operation are provided in the Supplementary Material 6 and the results of the Monte Carlo analysis after 1000 realizations are shown in figures 5.c to 5.e. Each TBU coupling constant is modeled by means of a Gaussian random variable centered at a mean value corresponding to its ideal setting and featuring a typical [41,42] value for the standard deviation $\sigma_K$= 1% accounting for random fluctuations around the mean. It should be noted here that the Monte Carlo analysis considers both the spatial and spectral behavior of the mesh although here only the transfer function $h_{36,15}$ is displayed so the input vector to the waveguide mesh is, in this case $I=(i_1,i_2...i_{40})$ where $i_k=0$ unless $k=16$ and $i_{16}=1$. Again, each realization takes only a few seconds, as it is based on an analytic method. The results provide very useful insight regarding relevant performance parameters such as the filter's extinction ratio, insertion losses and passband ripple, which can be displayed in terms of histograms from which averages and standard deviation values can be extracted. Our method allows not only for the Monte Carlo characterization of feedforward multiport interferometers in a similar way as that reported elsewhere [41,42], but also for complex feedforward and feed-backward configurations as shown by this example. We stress again that all realizations rely on analytic recursive expressions.

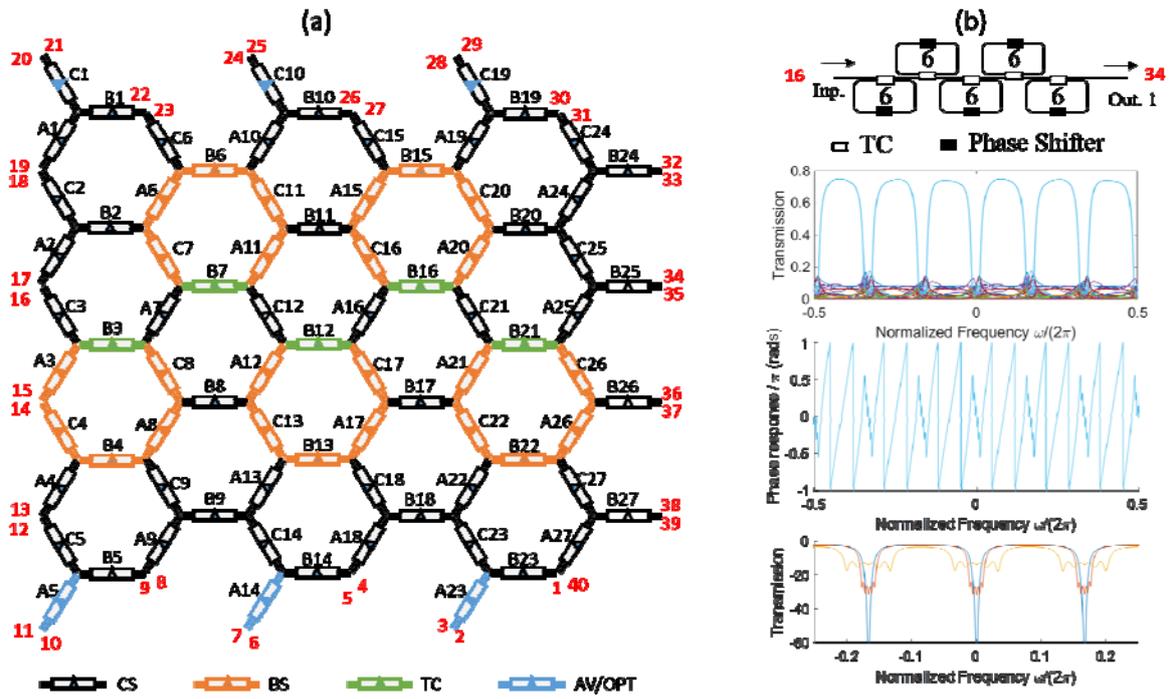

**Fig. 4** Scalable Analysis Method application to full spectral analysis of a waveguide mesh implementing a feedforward/feedbackward SCISSOR filter composed of 5 cascaded ring resonators with the same cavity length (6 BULs). (a) Mesh architecture and configuration for implementing the SCISSOR filter. (b) Equivalent circuit layouts with indication of the input and output ports in red ink (upper). Moduli of the 40 transfer functions obtained after multiplying the 40x40 matrix of spectral transfer functions by the input vector $I=(i_1,i_2,...i_{40})$ where $i_k=0, k\neq 16$ and $i_{16}=1$ for the case where $K=0.2$ and $\phi_1=\phi_2=\phi_3=\phi_4=\phi_5=0$ (upper trace). Phase response of $h_{34,16}$ (intermediate trace) and Spectral response $h_{34,16}$ (lower trace) for two different cases where the SCISSOR parameters are changed. Case 1: $K=0.2$ in all the rings and ring resonances are slightly detuned, $\phi_1=-0.12$, $\phi_2=-0.06$, $\phi_3=0$, $\phi_4=0.06$, $\phi_5=0.12$, to reduce the filter bandpass and main to secondary sidelobes. Second case coupling constants are apodized $K_1=0.39$, $K_2=0.47$, $K_3=0.55$, $K_4=0.63$, $K_5=0.71$, and the ring resonances strongly detuned $\phi_1=-0.4$, $\phi_2=-0.2$, $\phi_3=0$, $\phi_4=0.2$, $\phi_5=0.4$.

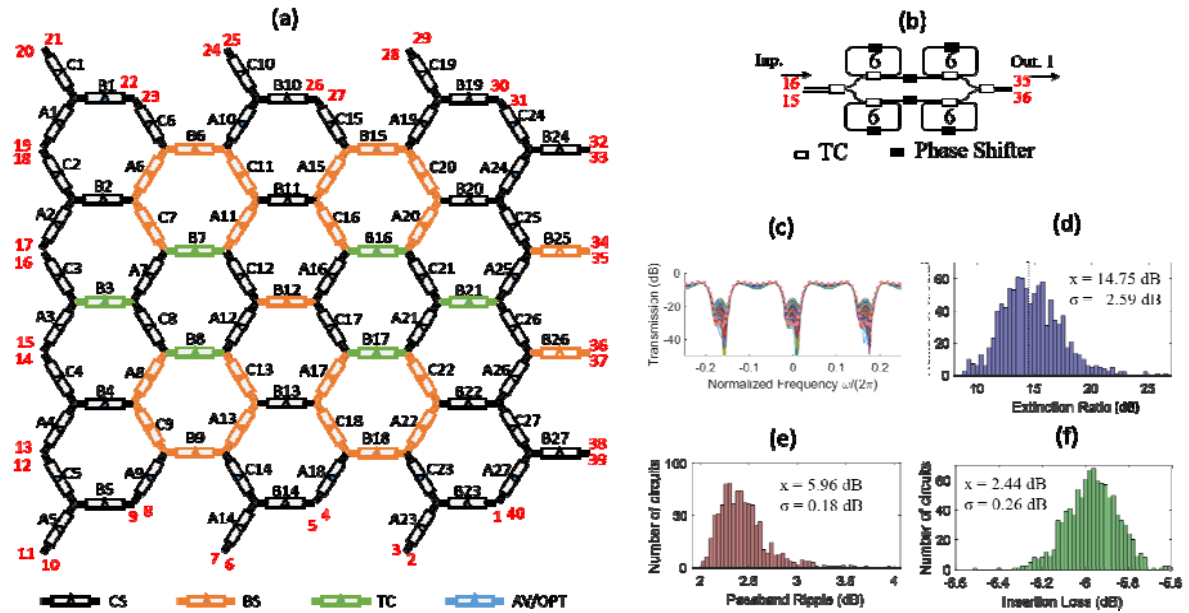

**Fig. 5 |** Scalable Analysis Method application to multiparameter error analysis of a waveguide mesh implementing a feedforward/feedbackward double ring loaded MZI with cavity lengths (6 BULs). (a) Mesh architecture and configuration for implementing the double ring loaded MZI. (b) Equivalent circuit layout with indication of the input and output ports in red ink. (c)-(f) Results of the Monte Carlo analysis of the spectral transfer function $h_{35,16}$ after 1000 realizations, where Each TBU coupling constant is modelled by means of a Gaussian random variable centered at a mean value corresponding to its ideal setting and featuring a standard deviation $\sigma_K=0.1$ accounting for random fluctuations around the mean. (c) Spectral transfer function realizations, (d) Filter Extinction ratio statistics, (e) Filter passband ripple statistics, (e) Filter insertion loss statistics

## D. Circuit Performance Optimization

A distinctive feature of the method proposed here is that it provides information related to all the possible signal paths and transfer functions established between input and output ports of the mesh once it has been programmed to implement a given circuit or a simultaneous group of circuits. The mesh is then divided into two parts. One part corresponds to the elements employed to implement the circuits while the other is composed of the elements that are not needed. This second part of the mesh can either be left as it stands or, more interestingly, it can be employed to improve the performance of the programmed circuits by establishing connections to drain possible sources of crosstalk between the implemented circuits to subsidiary ports. Since the method provides the complete information regarding the waveguide mesh, it can be readily employed to test the impact of the programming of non-essential parts of the mesh in order to optimize its operation. To our knowledge no other method reported so far is able to provide this feature. To illustrate this concept, we consider now an example where the waveguide mesh is programmed to implement two simple circuits; a three cavity CROW device [43] and a simple MZI filter. Figure 6.a. depicts the waveguide mesh programming (parameters are provided in the Supplementary Material 6) to implement the circuit layouts shown in Figure 6.b. When both circuits are running in parallel, signal coming from port 17 can partially leak and appear in ports 38 and 39. The crosstalk contributions are obtained from the scattering matrix as $h_{38,17}$, $h_{39,17}$, $h_{23,2}$, $h_{30,2}$, respectively. Moreover, the signal leaking impacts on the desired individual responses $h_{30,17}$, $h_{23,17}$, $h_{38,2}$, $h_{39,2}$. This is illustrated in Fig. 6.c. for both the CROW transmission, and reflection responses as well as for both outputs of the MZI by means a Monte Carlo with 1000 realizations and a TBU coupling factor standard deviation ($\sigma_K$) equal to 1 %. The analysis results predict crosstalk levels with average figures in the range of -49.11 dB to -43.6 dBs for the relevant transfer functions of interest when the unused TBUs are assumed to be randomly biased (see Figure 6.e). Moreover, the performance of the circuits is visibly compromised.

By suitable biasing of the unused TBUs (see Supplementary Material 6) the leaking signals can be re-directed to drain ports for elimination. For example, we have configured the TBUs paced outside the marked regions in the mesh to achieve the routing of these un-desired crosstalk signals to achieve a crosstalk improvement of 20 dB as shown in Fig 6.f. We ran the test for standard deviation levels $\sigma_K$ ranging from 0.5 % to 2 % obtaining a considerable improvement for the crosstalk levels in all the relevant circuit transfer functions as shown in Fig. 6.g (for the test details, please refer to Methods). The optimized transfer functions are visibly more robust, as illustrated in Fig.6.d, relaxing the specifications of each TBU. In addition, the circuit performance could be further improved by monitoring a few outputs and reprogramming the mesh accordingly to maximize the leaked signal evacuation and the isolation between the simultaneously programmed circuits.

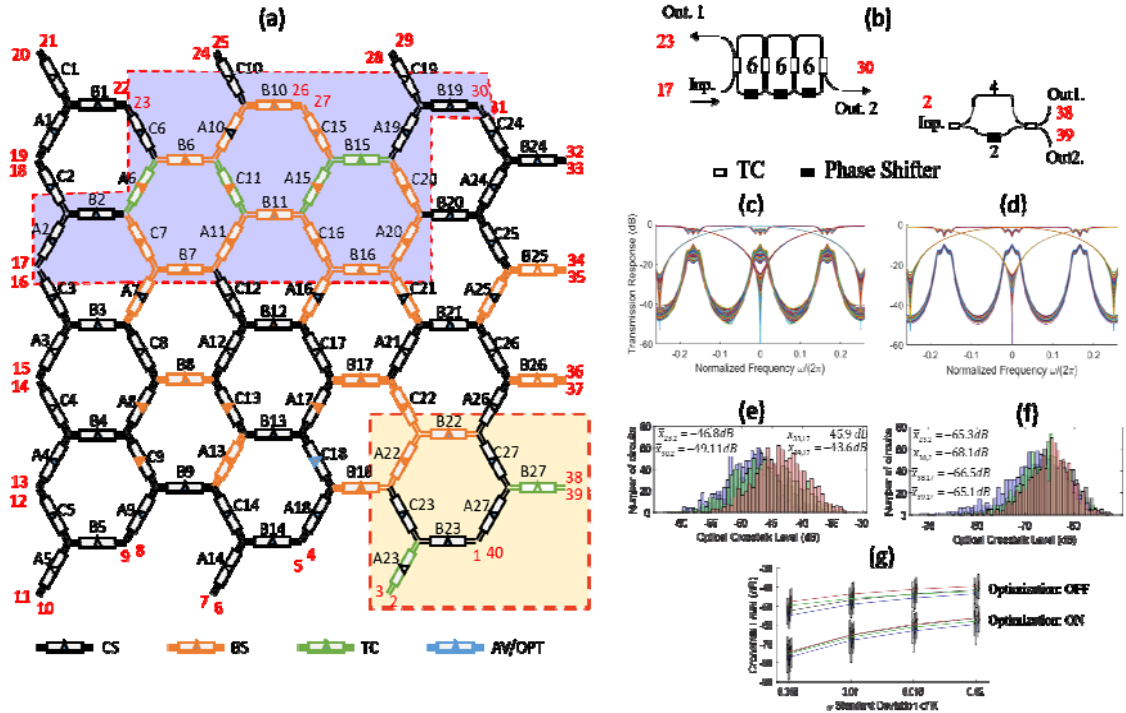

**Fig. 6** | Scalable Analysis Method application to circuit parameter optimization of a waveguide mesh implementing simultaneously a 3 stage CROW with cavity lengths (6 BULs) and a MZI. (a) Mesh architecture and configuration for implementing the two circuits. (b) Equivalent circuit layouts with indication of the input and output ports in red ink. (c) Monte Carlo results (1000 runs with $\sigma_K=0.01$) for the spectral transfer functions of the two circuits before optimization (d) Monte Carlo results for the spectral transfer functions of the two circuits after optimization. (e) Statistical results for crosstalk levels corresponding to transfer functions $h_{30,2}$, $h_{23,2}$, $h_{38,17}$ and $h_{39,17}$ before optimization and (f) after optimization. (g) Crosstalk levels for different values of standard deviation values of the coupling coefficients $\sigma_K$.

### E. Experimental verification

An experimental validation of the model is now provided. For this purpose we synthetized and measured different programmed circuits architectures in a seven-cell hexagonal waveguide mesh fabricated in silicon on insulator (see details in [8] ). By inserting the data of the experimental measurements carried out in [8], such as the insertion losses of the TBU, the Basic Unit Length, as well as the coupling and phase factors settings, we obtained a perfect matching after comparing the results with those provided by the model. As an example, Fig. 7 illustrates different experimental (solid) and simulated (dashed) traces for the tuning of both a ring resonator (a-b) and the transmission response of a CROW (c-d). In the first case the coupling of B7 was modified while tuning the phase inside the ring resonator. In the second case, the resonance of the upper cavity is slightly detuned and an imperfect bar state is synthetized at TBUA5.

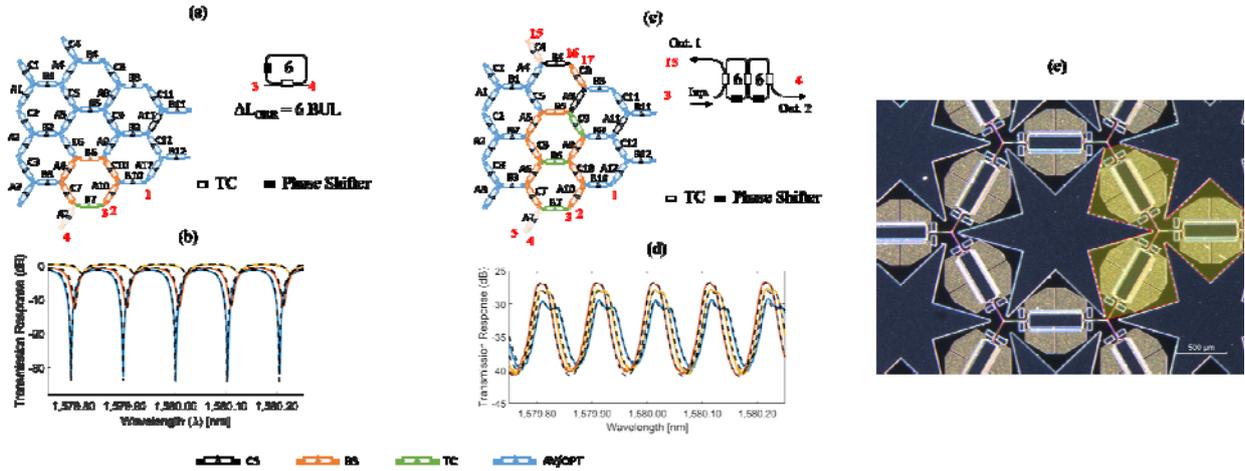

**Fig. 7.** Experimental validation of the model. (a) Programmed waveguide mesh and targeted circuit: Optical Ring Resonator of cavity length equal to 6-BULs. (b) Experimental (solid) and modeled (dashed) response for different circuits conditions of coupling and phase. (c) Programmed waveguide mesh and targeted circuit: CROW with two ring resonators slightly decoupled. (d) Experimental and modeled response for different phase detuning conditions. (e) Photograph of the fabricated device reported in [8].

## 4. METHODS

### A. Implementation

The method has been implemented using a script in MATLAB that takes into account the four different scenarios for increasing the waveguide mesh by one trilattice unit. For each case the relevant analytical equations derived in the Supplementary Material 1 are employed.

### B. Monte Carlo Simulations

For the Monte Carlo simulations we ran 1000 iterations. The insertion loss of the TBUs is kept fixed to a value of 0.2 dB unless specified. For each iteration, the value of each TBU coupling factor is selected from a normal distribution with average value corresponding to the originally programmed value for ideal operation and a standard deviation specified for each simulation (ranging from 0.5 % to 2 %). Unless specified, overall TBU phase variations are kept fixed to 0.

## 5. DISCUSSION AND SUMMARY

The proposed method can be extended to other waveguide mesh geometries (square, triangular) reported in the literature. The key point here is to identify the unitary building block upon which the mesh can be built (see the Supplementary Material 2 for a description of the unitary building blocks for other cell geometries). It can also be employed for the analysis of triangular and rectangular multibeam feedforward only interferometers. In this case it suffices to emulate the interferometer with the waveguide mesh and program the rest of the mesh to disable any potential signal route outside the physical topology of the interferometer. A final interesting application of the method to be explored in the near future is related to the use of the full scattering matrix to implement circuit state supervision by means of machine learning technique. This could alleviate the need for individual device monitoring by means of signal tapping or CLIPP based approaches [40, 45].

In summary, we have reported a scalable analytic method based on mathematical induction that renders the full scattering matrix of any 2D integrated photonic waveguide mesh circuit composed of an arbitrary number of hexagonal cells and which is easily programmable. The method not only provides all the desired input/output transfer functions, but also allows to design the unused regions of the waveguide mesh so they can be employed to manage undesired contributions from reflected signals and thus optimize the chip performance and furthermore, it allows to study all the input/output responses as the internal parameters of the TBUs are changed opening the path for error evaluation via Monte Carlo simulations. We believe that our results open the path to unblock an important bottleneck in the design of complex photonic circuits and will enable the fast analysis of large (LSI) and very large (VLSI) scale integrated multifunctional photonic circuits based on waveguide meshes.

**Funding**. The authors acknowledge financial support by the ERC ADG-2016 UMWP-Chip, the Generalitat Valenciana PROMETEO 2017/017 research excellency award, and the COST Action CA16220 EUIMWP through the FPI predoctoral funding scheme.

# Scalable Analysis for Arbitrary Photonic Integrated Waveguide Meshes: supplementary material

DANIEL PEREZ,[1] AND JOSE CAPMANY[1]*

[1] *ITEAM Research Institute, Universitat Politècnica de València, Valencia, 46022, Spain.*
*Corresponding author: dperez@iteam.upv.es*

This document provides supplementary information to "Scalable Analysis for Arbitrary Photonic Integrated Waveguide Meshes,"

.]

## 1. Inductive Method and scenarios

To illustrate the procedure of obtaining the scattering matrix of hexagonal waveguide meshes with an arbitrary number of Tunable Basic Units (TBUs), we describe here, for each scenario, the associated interconnection diagram and the resulting matrix after appending to a waveguide mesh composed of n-1 tri-lattices (or n-1 order mesh) an additional tri-lattice element to form an n order mesh.

***Scenario 0***: This is the simplest case and the starting point in the generation of a new mesh design. Here, only one out of the 6 ports of the new tri-lattice (*Latt* N) is connected to the n-1 order mesh. . As shown in Supplementary Figure 1.a the addition of the new tri-lattice increases the number of mesh ports by 4, and correspondingly, the number of rows and columns in the scattering matrix. The interconnection diagram shown in Supplementary Figure 1b illustrates the signal flow possibilities inside the n-1 order mesh and between this mesh and the newly added tri-lattice through the interface node *x=P*. This interconnection scheme defines a system of equations associated to node x that can be solved, rendering equations (1) that provide the matrix coefficients that characterize the new waveguide mesh ports.

The resulting matrix, shown in Supplementary Figure 1c, can be decomposed in four submatrices: The first submatrix (Submatrix 1) is related to the connections between the n-1 order mesh ports excluding the ones that will be interconnected to the new lattice (in this case port *P*). The second (Submatrix 2) relates the new inputs in *Latt* N to the output ports of mesh *n*-1. The third (Submatrix 3) relates the input ports in mesh *n*-1 with the new outputs pors in *Latt* N. Finally, Submatrix 4 describes the connections of the inputs/output ports of *Latt* N.

Submatrix 1 coefficients: $h_{s,r} = X = h_{s,r}^{N-1}$.

Submatrix 2 coefficients: $h_{s,(P,...,P+4)}^{N} = GB'$

Submatrix 3 coefficients: $h_{(P,...,P+4),r}^{N} = TS$,

Submatrix 4 coefficients: $h_{(P,...,P+4),(P,...,P+4)}^{N} = Th_{xx}G + IntCon$.

(S1)

where *IntCon* represents the internal connections given by the scattering matrix of the newly added trilattice *Latt* n.

***Scenario 1***: Here, the addition of the new tri-lattice (*Latt* N) increases the number of mesh ports by two but the number of complete hexagonal cells does not increase, as shown in Supplementary Figure 2.a. Supplementary Figures 2.b. and 2.c illustrate the associated interconnection diagram to be solved and the resulting matrix for the n order mesh respectively. In this case, the resulting equations are more complex since two interface nodes (*x=P-1* and *y=P*) are required.

Solving the system of equations related to nodes *x=P-1* and *y=P* renders equations (2) that provide the matrix coefficients that characterize the new waveguide mesh ports and the four submatrices:

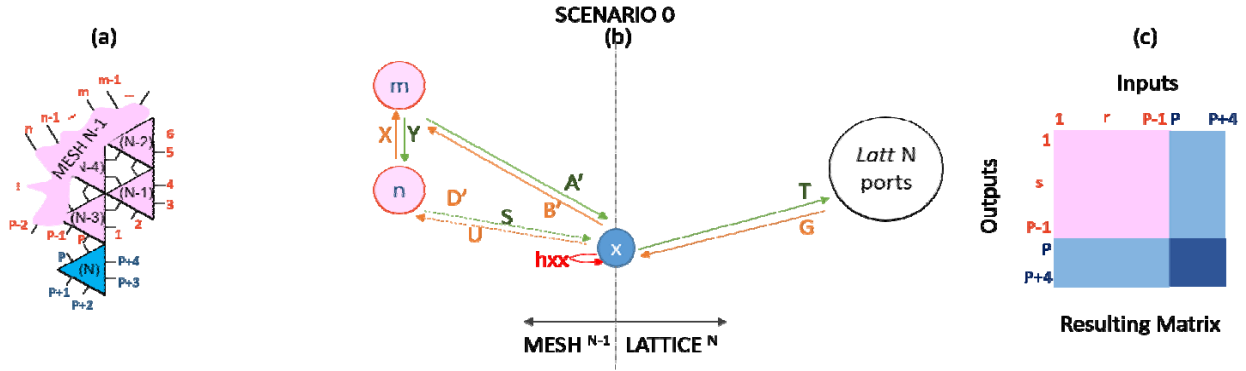

Fig. S1. **Scenario 0.** (a) Connection scheme of the additional trilattice with mesh *n-1*, (b) interconnection graph diagram with the labelled contributions, (c) resulting matrix sections. S0: x = P. Direct contributions inside *lattice N* ports are not included in the graph.

SM1  $h_{s,r} = X = h_{s,r}^{N-1}$.

SM2  $h_{s,(P-1,\ldots,P+2)} = B'G + DP$,

SM3  $h_{(P-1,\ldots,P+2),r} = OE' + TS$, (S2)

SM4  $h_{(P-1,\ldots,P+2),(P-1,\ldots,P+2)} = T(h_{xx}G + PM) + O(h_{yy}P + GN)$.

*Scenario 2*: Here, the addition of the new tri-lattice increases the number of ports by two and the number of complete hexagonal cells by one as shown in Supplementary Figure 3.a. In this case, the signal flow diagram in Supplementary figure 3.b includes the possibility of recirculation between the interfacing nodes x=P-1 and y=P and the newly added trilattice unit (*Latt* N) as shown by connections V and W. The procedure is similar to the two previous cases by solving the system of equations associated to the nodes y and x.

Solving the system of equations related to nodes *x=P-1* and *y=P* renders equations (3) that provide the matrix coefficients that characterize the new waveguide mesh ports and the four submatrices:

SM1  $h_{s,r} = X + \dfrac{\begin{bmatrix} DW\left[h_{xx}^{(N-1)}VE' + (1-VN)S\right] + \\ B'V\left[h_{yy}^{(N-1)}WS + (1-MW)E'\right] \end{bmatrix}}{(1-VN)(1-MW) - h_{xx}^{(N-1)}h_{yy}^{(N-1)}VW}$,

$h_{s,(P-1,\ldots,P+2)} = UG + F'P +$

SM2  $+ \dfrac{\begin{pmatrix} F'W\left[h_{xx}^{(N-1)}(G + VPh_{yy}^{(N-1)}) + (1-VN)MP\right] + \\ UV\left[h_{yy}^{(N-1)}(P + WGh_{xx}^{(N-1)}) + (1-MW)NG\right] \end{pmatrix}}{(1-VN)(1-MW) - h_{xx}^{(N-1)}h_{yy}^{(N-1)}VW}$, (S3)

SM3  $h_{(P-1,\ldots,P+2),r} = \dfrac{\begin{pmatrix} O\left[h_{yy}^{(N-1)}WS + (1-MW)E'\right] + \\ T\left[h_{xx}^{(N-1)}VF + (1-NV)S\right] \end{pmatrix}}{(1-VN)(1-MW) - h_{xx}^{(N-1)}h_{yy}^{(N-1)}VW}$,

$h_{(P-1,\ldots,P+2),(P-1,\ldots,P+2)} =$

SM4  $= \dfrac{\begin{pmatrix} O\left[h_{yy}^{(N-1)}P + h_{yy}^{(N-1)}Wh_{xx}^{(N-1)}G + (1-MW)NG\right] \\ + T\left[h_{xx}^{(N-1)}G + h_{yy}^{(N-1)}Vh_{xx}^{(N-1)}P + (1-VN)MP\right] \end{pmatrix}}{(1-VN)(1-MW) - h_{xx}^{(N-1)}h_{yy}^{(N-1)}VW}$

$+ IntCont$.

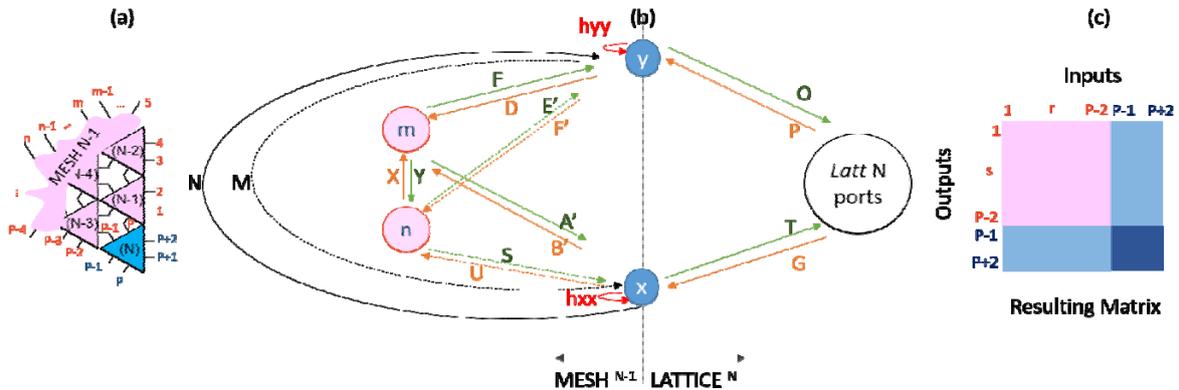

Fig. S2. **Scenario 1.** (a) Connection scheme with mesh *n-1*, (b) interconnection graph diagram with the labelled contributions, (c) resulting matrix sections. S1: x = P-1, y = P. Direct contribution inside *lattice N* ports is not included in the graph.

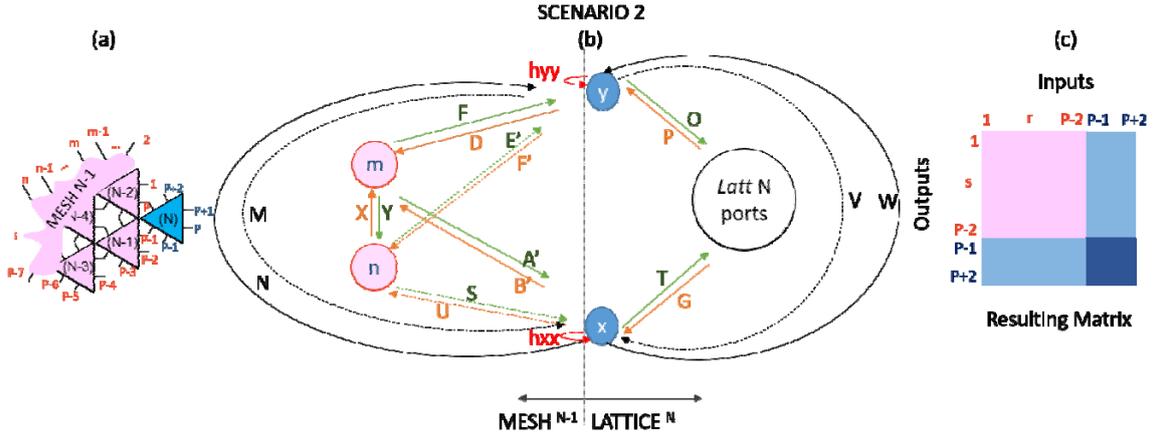

Fig. S3. Scenario 2. (a) Connection scheme with mesh *n*-1, (b) interconnection graph diagram with the labelled contributions, (c) resulting matrix sections. x = P-1, y = P. Direct contribution inside *lattice N* ports is not included in the graph.

**Scenario 3**: In this case as shown in Supplementary Figure 4.a, the addition the new tri-lattice does not increase the number of ports, since it connects three ports to the previous mesh and the number of complete cells is increased by one. Here, the interconnection diagram involves three interfacing nodes x, y, z, depicted in Supplementary Figure 4b. The procedure to obtain the coefficients of the different submatrices is similar to the three previous cases. Final expressions are included in (4).

$$z4 = \frac{H + CH_{xx}G + CMP + LH_{yy}P + LNH(b,tri\_lett)}{(1-BC-JL)}$$

$$h_{s,(P-1,...,P+2)} = D(P + Kz_1) + Rz_4 + B'(G + Ez_1)$$
$$\chi_1 = NE + HyyK,$$
$$\chi_2 = MK + HxxE,$$
$$\chi_3 = 1 - IK - FE,$$

$$\alpha_1 = (1-BC)\chi_1 + JC\chi_2;$$
$$\alpha_2 = (1-BC)\chi_3 - CH_{zz}\chi_2;$$

$$\beta_1 = (1-BC)E' + JCS;$$
$$\beta_2 = -(1-BC)C' - CH_{zz}S;$$

**SM1**

$$\xi_2 = C(MK + H_{xx}E) + L(NE + H_{yy}K),$$
$$\xi_1 = 1 - FE - IK - H_{zz}\xi_2,$$

$$z1 = \frac{\begin{pmatrix} C'(1-BC-JL)+ \\ (SC+E'L)Hzz \end{pmatrix}}{\begin{pmatrix} eps1(1-BC-JL)- \\ (BC+JL)Hzz\xi_2 \end{pmatrix}},$$

$$z4 = \frac{(CS + LE' + \xi_2 z_1)}{(1 - C.*B - L.*J)},$$

$$h_{s,r} = X + (DK + B'E)z_1 + Rz_4;$$

**SM2**

$$\xi_2 = C(MK + HxxE) + L(NE + HyyK)$$
$$\xi_1 = 1 - FE - IK - H_{zz}\xi_2$$

$$z1 = \frac{\begin{pmatrix} (1-BC-JL)(IP + FG) + \\ Hzz\begin{pmatrix} H + CHxxG + CMP \\ +LHyyP + LNG \end{pmatrix} \end{pmatrix}}{\begin{pmatrix} \xi_1(1-BC-JL)- \\ (BC+JL)H_{zz}\xi_2 \end{pmatrix}}$$

**SM3:**

$$z_5 = \frac{(1-BC-JL)\beta_2 - HzzL\beta 1}{(H_{zz}L\alpha_1 - (1-BC-JL)\alpha_2)},$$

$$y_3 = \frac{\beta 1 + \alpha_1 z_5}{(1 - BC - JL)},$$

$$x_3 = (S + BLy_3 + \chi_2 z_5)./(1-BC);$$

$$h_{(P-1,...,P+2),r} = Oy_3 + Az_5 + Tx_3,$$

(S4)

**SM4**

$$\chi_1 = NE + H_{yy}K;$$
$$\chi_2 = MK + H_{xx}E;$$
$$\chi_3 = 1 - IK - FE;$$

$$\alpha 1 = -\chi_2.*(Hyy.*P + J.*H + N.*G) + \ldots$$
$$\chi_1.*(Hxx.*G + B.*H + M.*P);$$

$$\alpha 2 = (1 - B.*C).*\chi_1 + J.*C.*\chi_2;$$

$$\alpha 3 = -\chi_3.*(Hxx.*G + B.*H + M.*P) - \ldots$$
$$\chi_2.*(Hzz.*H + I.*P + F.*G);$$

$$\alpha 4 = \chi_2 CHzz - \chi_3(1 - BC);$$

$$\beta 1 = (1 - JL)\chi_2 + BL\chi_1,$$
$$\beta 2 = \chi_3 BL + \chi_2 H_{zz}L,$$

$$y3 = (\alpha_3.*\alpha_2 - \alpha_1.*\alpha_4)./(\alpha_4.*\beta_1 + \alpha_2.*\beta_2);$$

$$x3 = (-y_3.*\beta 2 + \alpha_3)./\alpha_4;$$

$$z5 = \begin{pmatrix} y_3.*(1-JL) \\ -H_{yy}P - JH - JCx_3 - NG \end{pmatrix}./\chi_1;$$

or

$$z5 = \begin{pmatrix} H_{zz}H + IP + FG + \\ H_{zz}Cx_3 + HzzLy_3 \end{pmatrix}./\chi_3;$$

$$h_{(P-1,\ldots,P+2),(P-1,\ldots,P+2)} =$$
$$= Oy_3 + Az_5 + Tx_3 + IntCont.$$

## 2. Procedure to develop the model for different waveguide mesh topologies.

Waveguide meshes can be implemented using different unit cell geometries [1],[2] (square, hexagonal, triangular, etc.). These have been benchmarked against a series for photonic figures of merit and the hexagonal topology features the best performance in terms of most of them [1]. Nevertheless, it might be advisable for other reasons to implement the mesh following alternative unit cell geometries. the inductive method described here can also be applied in those cases to obtain the overall waveguide mesh scattering matrix provided that a suitable building block is identified. Supplementary Figure 5 highlights different building block alternatives for each of the main mesh geometries. Depending on the waveguide mesh topology and the chosen TBU, we can straightforwardly identify the different connection scenarios and proceed to solve their associated system of equations in the same way as is done in this paper for the tri-lattice and the hexagonal topology.

## 3. Further Examples on Single Wavelength Analysis

One of the main benefits of the method proposed resides in its powerful and flexible behavior. To further illustrate its versatility, here we include more examples where multiple 2x2 and 4x4 linear transformations are implemented by more than one circuit programmed over the same waveguide mesh in the case of single-wavelength operation. Moreover, we test the flexibility of the method to characterize the performance of waveguide meshes under realistic conditions by showing how it can be employed to evaluate the impact of having non-ideal components.

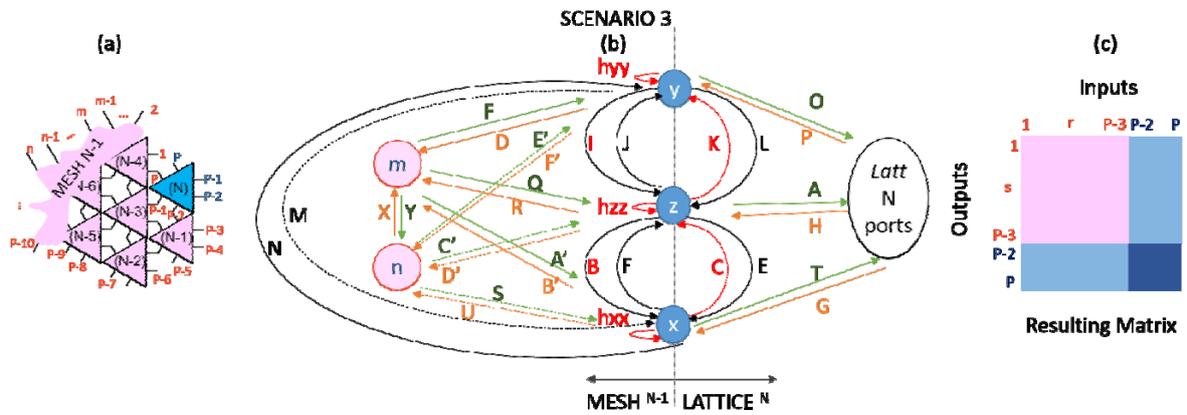

Fig. S4. Scenario 3. (a) Connection scheme with mesh *n*-1, (b) interconnection graph diagram with the labelled contributions, (c) resulting matrix sections. x = P-2, y = P-1, z = P. Direct contribution inside *lattice N* ports is not included in the graph.

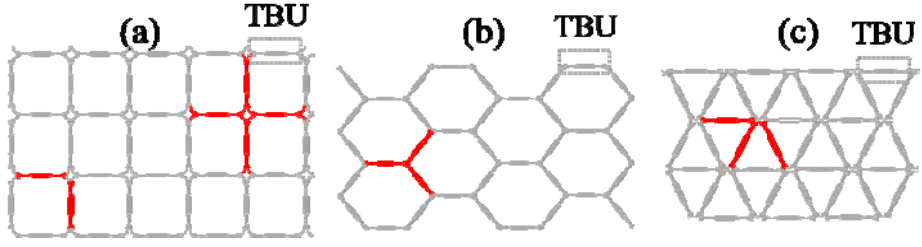

Fig. S**5**. Possible building blocks for the implementation of the inductive method described in the paper in different waveguide mesh cell topologies. The highlighted structures identify unitary elements upon which the given mesh topology can be constructed by suitable addition of them. (a) Square waveguide mesh, (b) Hexagonal waveguide mesh, (c) Triangular waveguide mesh.

**Linear transformation examples**

A wide variety of signal processing operations involve mode transformations, which can be described in terms of multiple input/multiple output linear optics transformations given by an N×N unitary matrix U, [3,4]. These include, among others, switching and broadcasting, mode combiners and splitters, and quantum logic gates. We programmed the waveguide mesh to demonstrate several 2x2, 3×3 and 4×4 linear transformations. These are relevant examples of signal processing tasks that are needed in different applications and the results are shown in Supplementary Figs. 6-8.

*2x2 Transformations (Hadamard, Pauli-y, z)*: Supplementary Figure 6a shows an example of three simultaneously programmed 2×2 transformations. These matrices, specified in (4), are the Y-, Z- Pauli gates and the Hadamard matrix. Supplementary Fig. 6a illustrates the configuration of the considered waveguide mesh. First, the circuit location is chosen for the three transformations (coloured background). Then, the access to the structures is provided by properly configuring the TBUs. The configuration of the phase and coupling factor of each TBU concerning the interferometric part is provided by the adaptation to the hexagonal waveguide mesh [5] of the original rectangular interferometer transformation, [4]. The resulting configuration coefficients are included in Supplementary Table 6.

$$H_{P-y} = \begin{pmatrix} 0 & -i \\ i & 0 \end{pmatrix}, \quad H_{Had} = \begin{pmatrix} 1 & 1 \\ 1 & -1 \end{pmatrix}, \quad H_{P-z} = \begin{pmatrix} 1 & 0 \\ 0 & -1 \end{pmatrix}. \quad \textbf{(S5)}$$

Supplementary Figure 6b illustrates the targeted circuit layouts and the access ports in the waveguide mesh (red ink). Supplementary Figure 6c. illustrates the computed amplitude and phase response of the overall mesh at the centre wavelength. A light-dashed line surrounds the targeted ports, resulting in a perfect implementation of the desired transformations and the additional undesired input/output port relationships that are enabled by the waveguide mesh programming.

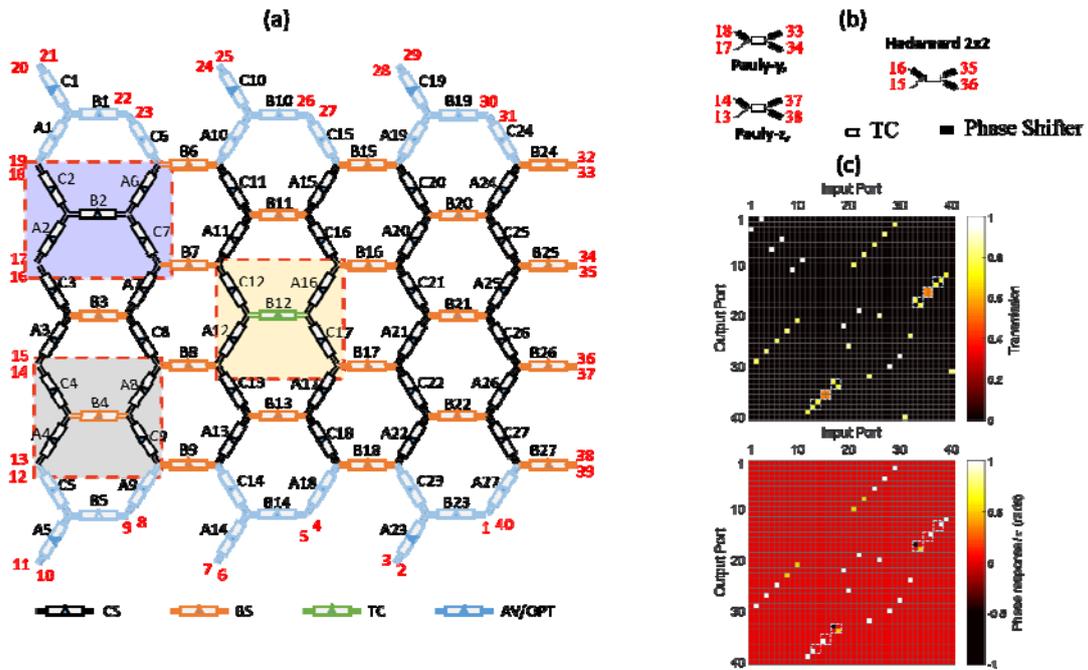

Fig. S6 **Scalable Analysis Method application to single wavelength operation of waveguide mesh configuration for universal linear interferometers.** (a) Mesh architecture and configuration for simultaneously implementing three 2x2 linear transformations. (b) Equivalent circuit layouts with indication of the input and output ports in red ink. (c) Moduli and phases of all the 40x40 matrix coefficients when the 2x2 transformations are programmed to implement a Pauli-Y, Hadamard and Pauli-Z, respectively.

***4x4 Transformations DFT & CNOT***: Here, we increase the complexity of the circuits by programming a larger size transformation. Supplementary Fig. 7a illustrates the implementation example of a linear transformation of a Controlled-NOT gate and simultaneously, a Discrete Fourier Transform (DFT) operator, both described by unitary 4×4 matrices, as specified in (5). The coupling and phase coefficients of each TBU are computed and specified in Supplementary Table 7. We can program the mesh to implement these gates in a very compact layout. Again, the results illustrated in Supplementary Fig.7c match with the targeted values of amplitude and phase and again, the 40x40 waveguide mesh scattering matrix provides information related to the additional undesired input/output port relationships that are enabled by the waveguide mesh programming.

$$H_{DFT4} = \frac{1}{2}\begin{pmatrix} 1 & 1 & 1 & 1 \\ 1 & -i & -1 & i \\ 1 & -1 & 1 & -1 \\ 1 & i & -1 & -i \end{pmatrix}, \quad H_{CNOT} = \begin{pmatrix} 1 & 0 & 0 & 0 \\ 0 & 1 & 0 & 0 \\ 0 & 0 & 0 & 1 \\ 0 & 0 & 1 & 0 \end{pmatrix}. \tag{S5}$$

***Non-unitary Linear transformations***: The Single Value Decomposition (SVD) can be used to decompose a non-unitary linear operator into two unitary operators and an intermediate diagonal matrix, [6]. For example, a linear transformation characterizes the operation of an optical hybrid used in coherent detection. This functionality can be described by a two by two operator described in (6). Then, the application of SVD results in three- matrices U, S, V' in (7), that satisfy $H_{Hyb}=USV'$.

$$H_{Hybrid} = \begin{pmatrix} 1 & 1 \\ 1 & -j \end{pmatrix} \tag{S6}$$

$$U = \frac{1}{\sqrt{2}}\begin{pmatrix} e^{-j0.875\pi} & e^{-j0.375\pi} \\ e^{j0.875\pi} & e^{j0.375\pi} \end{pmatrix},$$

$$S = \begin{pmatrix} 1 & 0 \\ 0 & 1 \\ 0 & 0 \\ 0 & 0 \end{pmatrix}, \tag{S7}$$

$$V' = \frac{1}{\sqrt{2}}\begin{pmatrix} -1 & 1 \\ e^{\frac{j3\pi}{4}} & e^{\frac{j3\pi}{4}} \end{pmatrix}.$$

For its implementation using the hexagonal waveguide mesh, we program the three matrices in a cascaded configuration V'-S-U as illustrated in the Supplementary Figure 8. The values of each TBU are provided by Supplementary Table 8. The SVD decomposition applied to the 2x4 Hybrid transformation in the main document is specified in (8), together with the corresponding values in Supplementary Table 1.

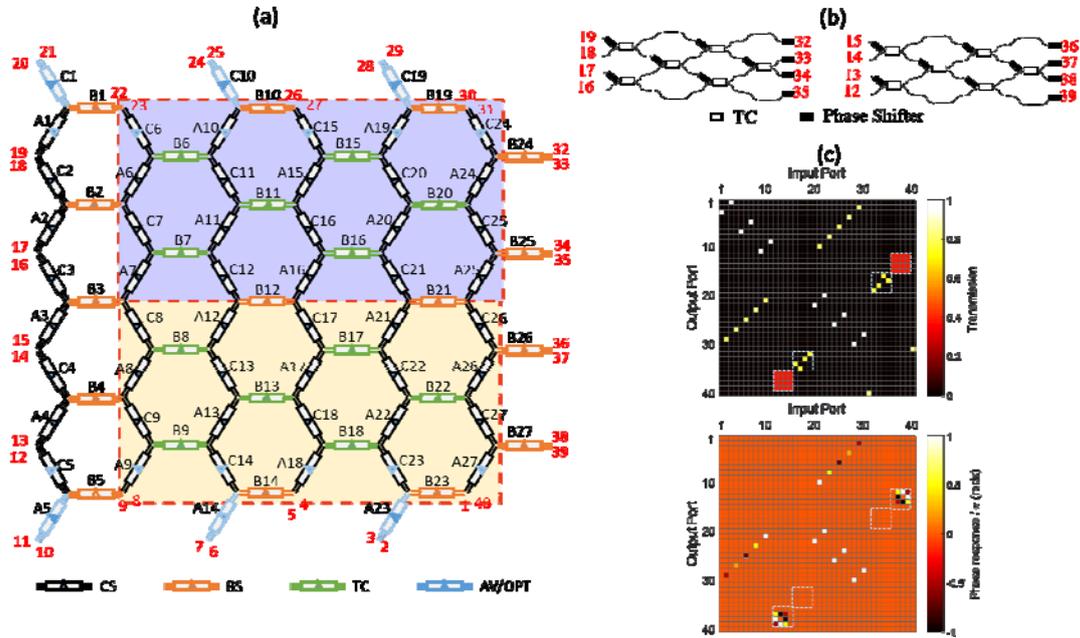

Fig. S7. **Scalable Analysis Method application to single wavelength operation of waveguide mesh configuration for universal linear interferometers.** (a) Mesh architecture and configuration for simultaneously implementing two 4x4 linear transformations. (b) Equivalent circuit layouts with indication of the input and output ports in red ink. (c) Moduli and phases of all the 40x40 matrix coefficients when the 2x2 transformations are programmed to implement a CNOT (Blue background) and Discrete Fourier Transform operator (Ochre background).

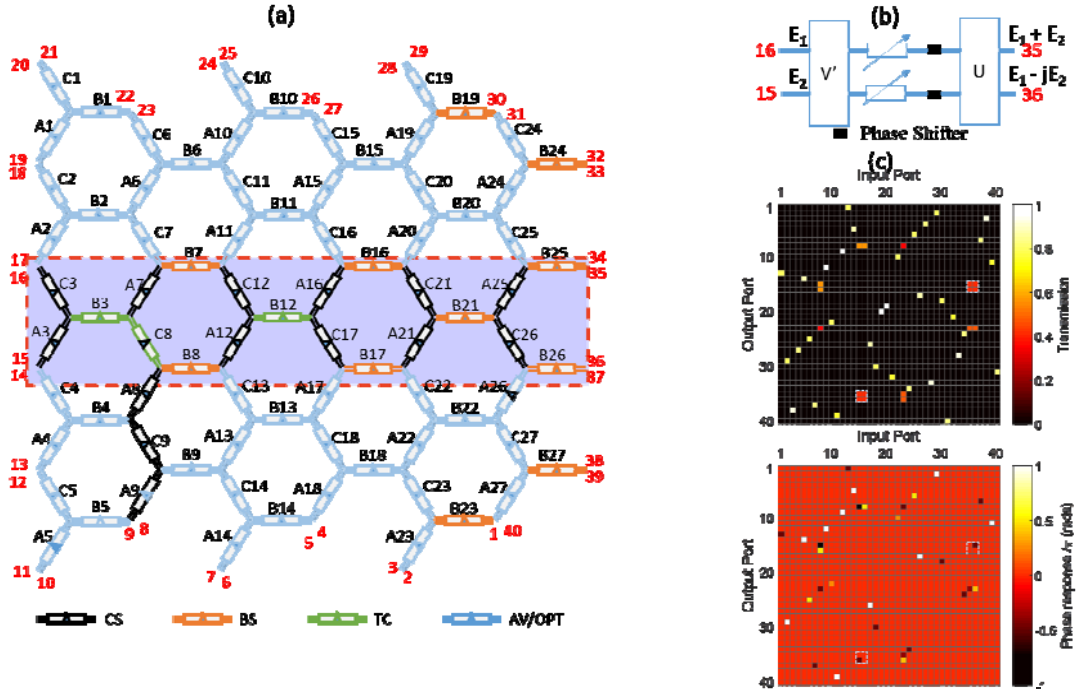

Fig. S8 **Scalable Analysis Method application to single wavelength operation of waveguide mesh configuration for universal linear interferometers.** (a) Mesh architecture and configuration for a 2x2 Hybrid operator described by cascading two 2x2 linear transformations and a phase and amplitude tapping array. (b) Equivalent circuit layout with indication of the input and output ports in red ink. (c) Moduli and phases of all the 40x40 matrix coefficients when the 2x2 transformation is programmed to implement an optical hybrid operator.

$$U = \frac{1}{2}\begin{pmatrix} e^{j\pi} & 1 & e^{j0.795\pi} & e^{-j0.795\pi} \\ e^{j\pi} & e^{j\pi} & e^{-j0.705\pi} & e^{j0.705\pi} \\ e^{j\pi} & e^{j\pi/2} & \sqrt{2}e^{j0.045\pi} & 0 \\ e^{j\pi} & e^{-j\pi/2} & 0 & \sqrt{2}e^{-j0.045\pi} \end{pmatrix},$$

$$S = \begin{pmatrix} 1 & 0 \\ 0 & 0.4142 \end{pmatrix},$$

$$V' = \begin{pmatrix} -1 & 0 \\ 0 & 1 \end{pmatrix}.$$

(S8)

*Analysis for non-ideal components*: The non-ideal behaviour of the TBUs leads to additional losses, scattering, and errors when setting the values for the phases and coupling factors. In the particular case of the implementation and analysis of linear transformations, we can use a simple model that assumes equal insertion loss for every TBU and normal coupling factor and phase drift distribution with standard deviations corresponding to $\sigma_K$ and $\sigma_\varphi$, respectively (See Supplementary Note 6 for TBU modelling information). We can quantify the fidelity of the transformation NxN $T_{exp}$ compared to the ideal/targeted transformation $T$ by using the following figure of merit, [4]:

$$F(T_{exp}, T) = \left| \frac{tr(T^\dagger T_{exp})}{\sqrt{N \cdot tr(T_{exp}^\dagger T_{exp})}} \right|^2 \quad \text{(S9)}$$

If $\sigma_K$ and $\sigma_\varphi$ are equal to 0. The fidelity of the hexagonal waveguide mesh remains equal to 1 (100%) independently of the circuit size and the loss per beam splitter. This is not the case for the triangular interferometer design [3] and the rectangular design [4]. The reason is that for the synthesis of linear transformations in the hexagonal waveguide mesh configuration, the number of TBU where the signal goes through is always equal for any possible lightpath. However, this might come at the cost of larger insertion loss.

Supplementary Figure 9 illustrates, as an example, the non-ideal component fidelity analysis for the two transformations (CNOT and DFT4) displayed in the example of Supplementary Figure 7. Here, we illustrate the test for two different $\sigma_K$ equal to 0.5% and 1%, whereas $\sigma_\varphi$ is varied between 0 and 0.5%.

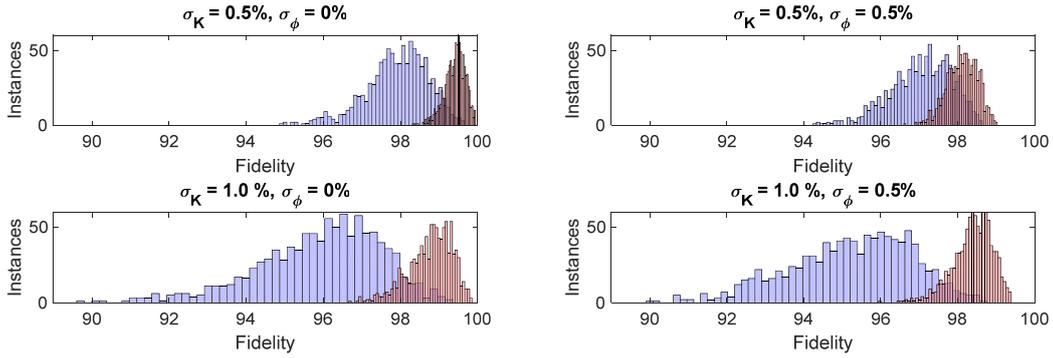

Fig. S9 . **MonteCarlo test to obtain the fidelity variations to non-ideal effects of the TBUs.** (Blue instances are related to CNOT transformation and red instances are related to DFT4 of Supplementary Figure 7.

The medians of the Fidelity distributions are:

$\sigma_K$ = 0.5%, $\sigma_\varphi$=0%: 98.00 %, 99.45 %,
      $\sigma_K$ = 0.5%, $\sigma_\varphi$=0.5%: 97.13 %, 98.14 %,
$\sigma_K$ = 1.0%, $\sigma_\varphi$=0%: 96.11 %, 98.86 %,
      $\sigma_K$ = 1.0%, $\sigma_\varphi$=0.5%: 95.34 %, 98.44 %,

Note that these results tend to confirm the findings reported by Burgwal and co-workers [7] in the sense that given a linear transformer topology, the fidelity is dependent on the application. Note as well that our method of analysis can include simultaneously the effect of errors in both the TBU power splitting ratio and phase shift and the TBU model could be substituted by any desired alternative.

## 4. Full spectral analysis

***Synthesis and analysis of photonic integrated circuits***: As mentioned in the main document, the more powerful and versatile characteristics of the analytic method are unleashed when using it for spectral characterization. Here, the wavelength (or frequency) dimension is added and truly spectral transfer functions are immediately provided in a few seconds. Moreover, the method can be readily applied in the analysis of complex setups resulting either from the simultaneous programming of multiple independent circuits over the mesh or from involved multistage resonant filters. We provide here two application examples to each case illustrated in Supplementary Figures 10 and 11 respectively. The case illustrated in Supplementary Figure 10 correspond to the simultaneous programming of three different ring cavities (of different cavity lengths) in the same waveguide mesh. The method generates a 40x40 matrix of transfer functions, each one spanning 1000 wavelengths. However, for practical purposes the only interesting transfer functions are $h_{22,20}$, $h_{7,8}$ and $h_{3,1}$, which can be recovered form the main system matrix by using an input vector given by $I = (i_1, i_2, \ldots i_{40})$ where $i_k = 0, k \neq \{3,8,20\}$ and $i_3 = i_8 = i_{20} = 1$. Note that the relevant transfer functions are exactly recovered and moreover, the remaining undesired contribution can also be retrieved (although they are not shown in the figure). In addition, the method allows to investigate the effect of changing individual phase shifters in the ring cavities.

Supplementary Fig. 11 illustrates the model application to a complex 2D resonant structure (in this case a three stage SCISSOR each one composed of a two cavity CROW. The schematic is depicted in Supplementary Figure 11b. Once configured by employing the coefficients in Supplementary Table 10, we obtained the transmission and reflection responses shown in Supplementary Fig. 11 c1. We can see the effect of switching off columns of CROWs by modifying the corresponding coupling coefficients to K = 0. The different traces for switching on 1, 2 and three second-order CROWs are illustrated in Supplementary Figure 11 c2. The method generates again the 40x40 matrix of transfer functions, each one spanning 1000 wavelengths. The interesting transfer functions are $h_{33,13}$ and $h_{32,13}$, which characterize the transmitted and reflected signals respectively and can be recovered form the main system matrix by using an input vector given by $I = (i_1, i_2, \ldots i_{40})$ where $i_k = 0, k \neq 13$ and $i_{13} = 1$. Again, the scalable method provides a fast and exact determination of the transfer functions even for this particularly involved structure where both longitudinal and lateral coupling and recirculations are allowed.

## 5. Relation between Overall TBU Coupling/Phase variation and internal Couplers variation in the balanced Mach-Zehnder configuration.

The Monte-Carlo test performed in this work consider the variation of the whole coupling and phase response of a generic TBU, rather than the variation in the inner couplers and phase-modulators in a balanced Mach-Zehnder configuration. This decision is due to the fact that the TBU architecture is not constrained to the MZI configuration and several alternatives can be employed. To compare the variability in both approaches, we have performed series of 1000-instances Monte-Carlo Test of the MZI-based TBU architecture considering different standard deviation of the inner couplers. These are defined by two independent normal distributions with mean 0.5 (blue-trace). For comparison, we also consider the effect of having the same coefficient in the input and output coupler (red-trace). We show that the numbers employed during the paper of $\sigma_K$ equal to 0.5% and 1% are in good agreement to the ones achievable with the state-of-the-art fabrication and design techniques.

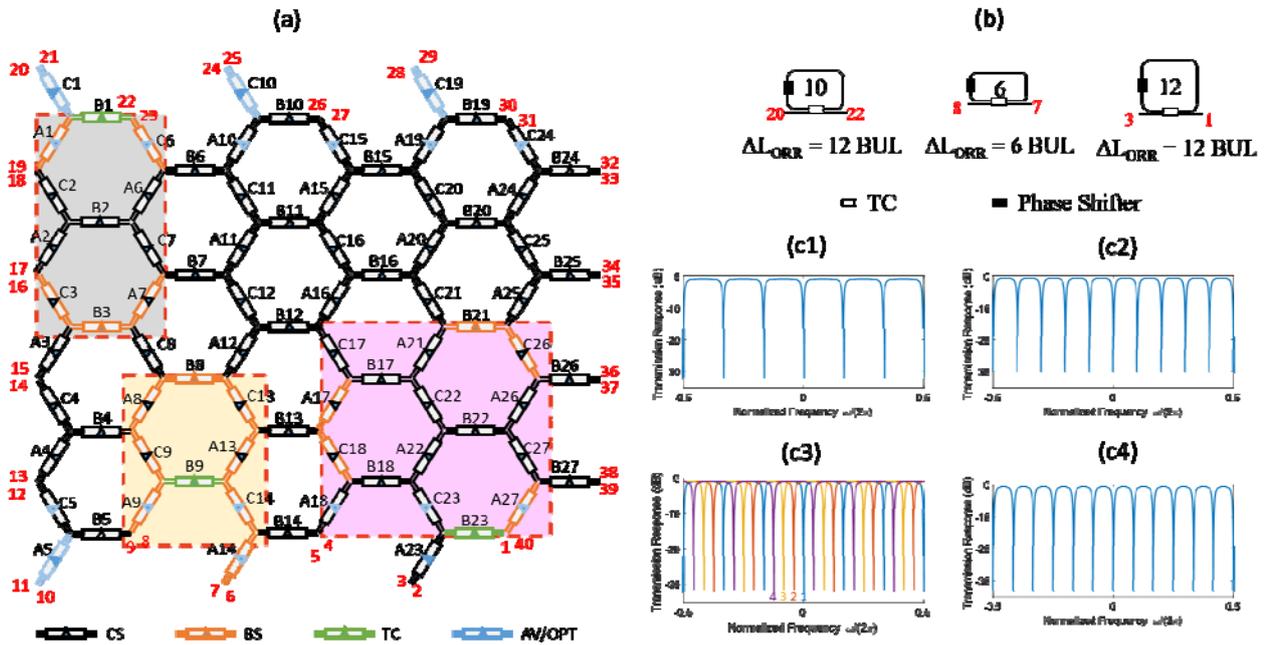

Fig. S10 **Scalable Analysis Method application to full spectral analysis of a waveguide mesh implementing three Optical Ring Resonator filters composed of cavity lengths equal to 6, 10, 12 TBUs** (a) Mesh architecture and configuration for simultaneously implementing the three filters. (b) Equivalent circuit layouts with indication of the input and output ports in red ink (upper). (c1) Transmission response of the 6-BUL ORR, (c2) Transmission Response of the 10-BUL ORR and (c4) the 12-BUL ORR. (c3) Tunability response of 6-BUL ORR for $\phi c_{13}$=0, 0.5, 1 and 1.5, respectively. The coupling factors are included in Supplementary Table 9.

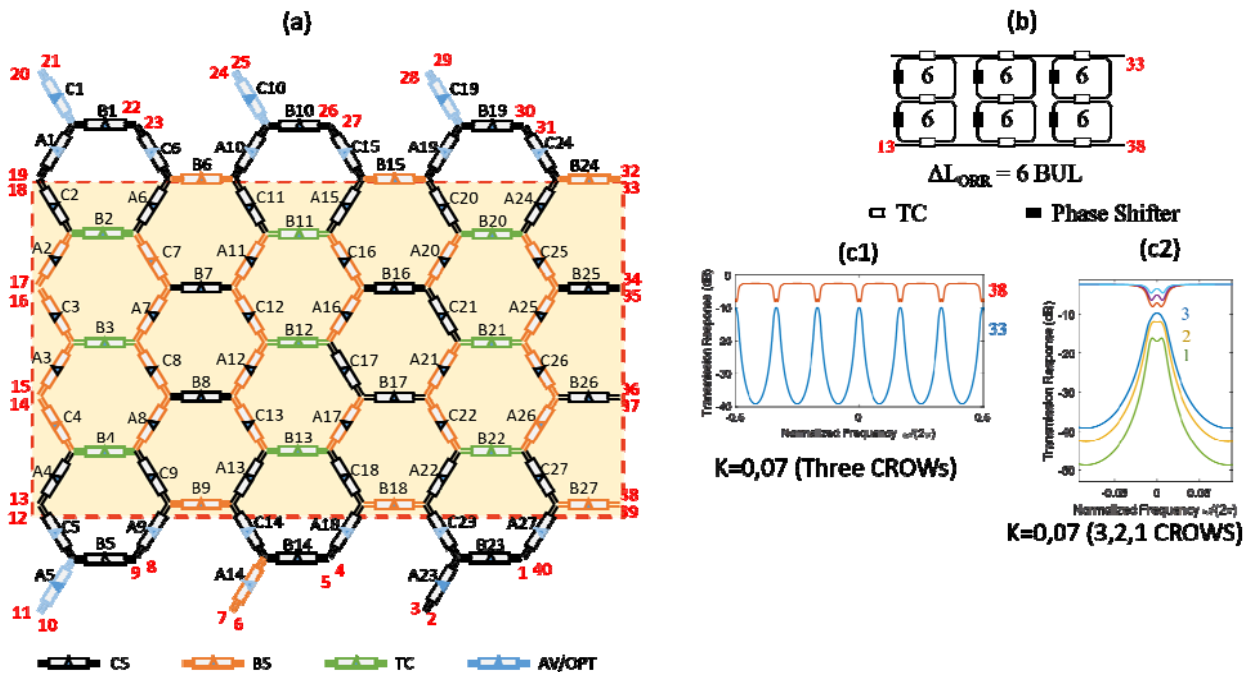

Fig. S 11. (a) Waveguide mesh configuration for the implementation of a three stage SCISSOR where each stage is composed of a second order CROW. Note that both lateral and longitudinal recirculations are allowed in this structure. (b) Circuit layout, (c1) Spectral Response (moduli) for equal value of the resonator coupling consntants K=0.07, (c2) Spectral Response (moduli) when switching off the third and both third and second CROW units.

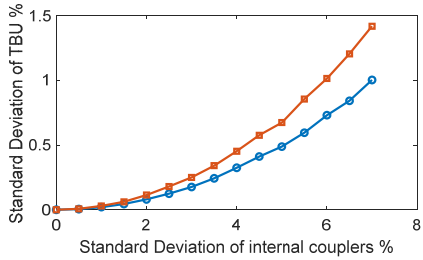

Fig. S12. Relation between the MZI-TBU approach inner couplers variation and the overall TBU coupling factor standard deviation. Each point is the mean of 1000-instances Monte-Carlo simulations where the inners couplers of the MZI-TBU distribution are independent (blue trace) and equal (red-trace).

The TBU simplified model employed in the simulations is therefore:

$$TBU = \alpha \begin{pmatrix} ie^{i\phi}\sqrt{1-K} \cdot e^{i\omega_n} & ie^{i\phi}\sqrt{K} \cdot e^{i\omega_n} \\ ie^{i\phi}\sqrt{K} \cdot e^{i\omega_n} & -ie^{i\phi}\sqrt{1-K} \cdot e^{i\omega_n} \end{pmatrix}, \quad \textbf{(S10)}$$

where $K$ is the coupling factor in power, $\phi$ is the overall phase term, $\omega_n$ is the normalised frequency (to the BUL) and $\alpha$ is the loss term. However, any model could be straightforwardly employed, including experimental data, or the balanced MZI response with the upper and lower phase terms as arguments.

## 6. Tables with the coupling and phase configuration for each programmed PIC in the document

Tables with coupling and phase values. For the implementation of linear transformations, we have employed the adaptation of the rectangular interferometer approach [4], to be programmed over a hexagonal waveguide mesh performed in [5].

**Supplementary Table1:** Coupling coefficients and phase coefficients normalized to π for the results showed in Fig. 3in the main document. DFT and a 2x4 Optical Hybrid respectively.

| TriLatt | KA | KB | KC | TA/pi | TB/pi | TC/pi |
|---|---|---|---|---|---|---|
| 1 | 1 | 0 | 1 | 0 | 0 | 0 |
| 2 | 1 | 0 | 1 | 0 | 0 | 1 |
| 3 | 1 | 0 | 1 | 0 | 0 | 0 |
| 4 | 1 | 0 | 1 | 0 | 0 | 1 |
| 5 | 1 | 0 | 1 | 0 | 1 | 0 |
| 6 | 1 | 0.5 | 1 | 0 | 0 | -1 |
| 7 | 1 | 0 | 1 | 0 | 1 | 0 |
| 8 | 1 | 0.5 | 1 | 0 | 0 | 1.5 |
| 9 | 1 | 0.5 | 1 | 0 | 0 | -0.4097 |
| 10 | 1 | 0 | 1 | 0 | 0 | 0 |
| 11 | 1 | 0.6667 | 1 | 0 | 0 | 1 |
| 12 | 1 | 0 | 1 | 0 | 0 | 0 |
| 13 | 1 | 0.6667 | 1 | 0 | 0 | -0.4548 |
| 14 | 1 | 0 | 1 | 0 | 1 | 0 |
| 15 | 1 | 0.5 | 1 | 0 | 0 | -0.5 |
| 16 | 1 | 0 | 1 | 0 | 1 | 0 |
| 17 | 1 | 0.75 | 1 | 0 | 0 | 0.0452 |
| 18 | 1 | 0.75 | 1 | 0 | 0 | 0 |
| 19 | 1 | 0 | 1 | -0.5 | 0 | 0 |
| 20 | 1 | 0 | 1 | -0.5 | 0 | 0.5 |
| 21 | 1 | 0 | 1 | 0 | 0 | 0 |
| 22 | 1 | 0.6667 | 1 | 0 | 0 | -1 |
| 23 | 1 | 0 | 1 | 0 | 1 | 0 |
| 24 | 1 | 0 | 1 | 0 | -0.5 | 0 |
| 25 | 1 | 0 | 1 | 0 | -0.5 | 0 |
| 26 | 1 | 0 | 1 | -1.2952 | 0 | -0.7952 |
| 27 | 1 | 0 | 1 | -2.0452 | 0 | -0.5452 |

**Supplementary Table 2:** Coupling coefficients and phase coefficients normalized to π for the results showed in Fig. 3 in the main document. Three-way beamsplitter and a 4x4 Hadamard matrix respectively.

| TriLatt | KA | KB | KC | TA/pi | TB/pi | TC/pi |
|---|---|---|---|---|---|---|
| 1 | 1 | 0 | 1 | 0 | 0 | 0 |
| 2 | 1 | 0 | 1 | 0 | 0 | 1 |
| 3 | 1 | 0 | 1 | 0 | 0 | 0 |
| 4 | 1 | 0 | 1 | 0 | 0 | 1 |
| 5 | 1 | 0 | 1 | 0 | 1 | 0 |
| 6 | 1 | 0.5 | 1 | 0 | 0 | -0.33334 |
| 7 | 1 | 0 | 1 | 0 | 1 | 0 |
| 8 | 1 | 0.5 | 1 | 0 | 0 | 0 |
| 9 | 1 | 0.5 | 1 | 0 | 0 | 0 |
| 10 | 1 | 0 | 1 | 0 | 0 | 0 |
| 11 | 1 | 0.6667 | 1 | 0 | 0 | -0.3334 |
| 12 | 1 | 0 | 1 | 0 | 0 | 0 |
| 13 | 1 | 1 | 1 | 0 | 0 | -1.1476 |
| 14 | 1 | 0 | 1 | 0 | 1 | 0 |
| 15 | 1 | 0.5 | 1 | 0 | 0 | 0.16667 |
| 16 | 1 | 0 | 1 | 0 | 1 | 0 |
| 17 | 1 | 0.5 | 1 | 0 | 0 | -1 |
| 18 | 1 | 0.5 | 1 | 0 | 0 | 0.1476 |
| 19 | 1 | 0 | 1 | -0.5 | 0 | 0 |
| 20 | 1 | 0 | 1 | 0.16667 | 0 | -0.16667 |
| 21 | 1 | 0 | 1 | 0 | 0 | 0 |
| 22 | 1 | 1 | 1 | 0 | 0 | 0.3185 |
| 23 | 1 | 0 | 1 | 0 | 1 | 0 |
| 24 | 1 | 0 | 1 | 0 | -0.5 | 0 |
| 25 | 1 | 0 | 1 | 0 | -0.5 | 0 |
| 26 | 1 | 0 | 1 | -1 | 0 | 0 |
| 27 | 1 | 0 | 1 | -2 | 0 | -1.3185 |

**Supplementary Table 3:** Coupling coefficients and phase coefficients normalized to π for the results showed in Fig. 4 in the main document. SCISSOR filter composed of 5 cascaded rings. Case 1: $K=0.2$ in all the rings and ring resonances are slightly detuned, $\varphi_1=-0.12$, $\varphi_2=-0.06$, $\varphi_3=0$, $\varphi_4=0.06$, $\varphi_5=-0.12$, to reduce the filter bandpass and main to secondary sidelobes. Second case coupling constants are apodized $K_1=0.39$, $K_2=0.47$, $K_3=0.55$, $K_4=0.63$, $K_5=0.71$, and the ring resonances strongly detuned $\phi_1=-0.4$, $\phi_2=-0.2$, $\phi_3=0$, $\phi_4=0.2$, $\phi_5=0.4$.

| TriLatt | KA | KB | KC | TA/pi | TB/pi | TC/pi |
|---|---|---|---|---|---|---|
| 1 | 1 | 1 | 1 | 0 | 0 | 0 |
| 2 | 1 | 1 | 1 | 0 | 0 | 0 |
| 3 | 0 | K1 | 1 | 0 | 0 | 0 |
| 4 | 1 | 0 | 0 | 0 | φ1 | 0 |
| 5 | 1 | 1 | 1 | 0 | 0 | 0 |
| 6 | 0 | 0 | 1 | 0 | φ2 | 0 |
| 7 | 1 | K2 | 0 | 0 | 0 | 0 |
| 8 | 0 | 1 | 0 | 0 | 0 | 0 |
| 9 | 1 | 1 | 1 | 0 | 0 | 0 |
| 10 | 1 | 1 | 1 | 0 | 0 | 0 |
| 11 | 0 | 1 | 0 | 0 | 0 | 0 |
| 12 | 0 | K3 | 1 | 0 | 0 | 0 |
| 13 | 1 | 0 | 0 | 0 | 0 | 0 |
| 14 | 1 | 1 | 1 | 0 | 0 | 0 |
| 15 | 0 | 0 | 1 | 0 | Φ4 | 0 |
| 16 | 1 | K4 | 0 | 0 | 0 | 0 |
| 17 | 0 | 1 | 0 | 0 | 0 | 0 |
| 18 | 1 | 1 | 1 | 0 | 0 | 0 |
| 19 | 1 | 1 | 1 | 0 | 0 | 0 |
| 20 | 0 | 1 | 0 | 0 | 0 | 0 |
| 21 | 0 | K5 | 1 | 0 | 0 | 0 |
| 22 | 1 | 0 | 0 | 0 | φ5 | 0 |
| 23 | 1 | 1 | 1 | 0 | 0 | 0 |
| 24 | 1 | 1 | 1 | 0 | 0 | 0 |
| 25 | 1 | 1 | 1 | 0 | -0.5 | 0 |
| 26 | 0 | 1 | 0 | 0 | 0.5 | 0 |
| 27 | 1 | 1 | 1 | 0 | 0 | 0 |

**Supplementary Table 4:** Coupling coefficients and phase coefficients normalized to π for the results showed in Fig. 5 in the main document. MZI loaded with four ORRs. The values in the table are considered as the average values for the normal distributions in the Monte-Carlo configurations.

| TriLatt | KA | KB | KC | TA/pi | TB/pi | TC/pi |
|---|---|---|---|---|---|---|
| 1 | 1 | 1 | 1 | 0 | 0 | 0 |
| 2 | 1 | 1 | 1 | 0 | 0 | 0 |
| 3 | 1 | 0.5 | 1 | 0 | 0 | 0 |
| 4 | 1 | 1 | 1 | 0 | 0 | 0 |
| 5 | 1 | 1 | 1 | 0 | 0 | 0 |
| 6 | 0 | 0 | 1 | 0 | 0.3819 | 0 |
| 7 | 1 | 0.59 | 0 | 0 | 0 | 0 |
| 8 | 0 | 0.59 | 1 | 0 | 0 | 0 |
| 9 | 1 | 0 | 0 | 0 | -0.382 | 0 |
| 10 | 1 | 1 | 1 | 0 | 0 | 0 |
| 11 | 0 | 1 | 0 | 0 | 0 | 0 |
| 12 | 1 | 0 | 1 | 0 | 0 | 0 |
| 13 | 0 | 1 | 0 | 0 | 0 | 0 |
| 14 | 1 | 1 | 1 | 0 | 0 | 0 |
| 15 | 0 | 0 | 1 | 0 | 0.581 | 0 |
| 16 | 1 | 0.999 | 0 | 0.112 | 0 | 0 |
| 17 | 0 | 0.999 | 1 | 0 | 0 | -0.112 |
| 18 | 1 | 0 | 0 | 0 | -0.581 | 0 |
| 19 | 1 | 1 | 1 | 0 | 0 | 0 |
| 20 | 0 | 1 | 0 | 0 | 0 | 0 |
| 21 | 1 | 0.5 | 1 | 0 | 0 | 0 |
| 22 | 0 | 1 | 0 | 0 | 0 | 0 |
| 23 | 1 | 1 | 1 | 0 | 0 | 0 |
| 24 | 1 | 1 | 1 | 0 | 0 | 0 |
| 25 | 1 | 0 | 1 | 0 | 0 | 0 |
| 26 | 1 | 0 | 1 | 0 | 0 | 0 |
| 27 | 1 | 1 | 1 | 0 | 0 | 0 |

**Supplementary Table 5:** Coupling coefficients and phase coefficients normalized to π for the results showed in Fig. 6 in the main document. Waveguide mesh implementing simultaneously a 3 stage CROW with cavity lengths (6 BULs) and a MZI. The values in the table are considered as the average values for the normal distributions in the Monte-Carlo configurations. The values in red are associated to the optimization implementation.

| TriLatt | KA | KB | KC | TA/pi | TB/pi | TC/pi |
|---|---|---|---|---|---|---|
| 1 | 1 | 1 | 1 | 0 | 0 | 0 |
| 2 | 1 | 1 | 1 | 0 | 0 | 0 |
| 3 | 1 | 1 | 1 | 0 | 0 | 0 |
| 4 | 1 | 1 | 1 | 0 | 0 | 0 |
| 5 | 1 | 1 | 1 | 0 | 0 | 0 |
| 6 | 0.15 | 0 | 1 | 0 | 0 | 0 |
| 7 | 0 | 0 | 0 | 0 | 0 | 0 |
| 8 | 1 | 0 | 1 | 0 | 0 | 0 |
| 9 | 1 | 1 | 1 | 0 | 0 | 0 |
| 10 | 0 | 0 | 1 | 0 | 0 | 0 |
| 11 | 0 | 0 | 0.2 | 0 | 0 | 0 |
| 12 | 1 | 1 | 1 | 0 | 0 | 0 |
| 13 | 0 | 1 | 1 | 0 | 0 | 0 |
| 14 | 1 | 1 | 1 | 0 | 0 | 0 |
| 15 | 0.2 | 0.15 | 0 | 0 | 0 | 0 |
| 16 | 0 | 0 | 0 | 0 | 0 | 0 |
| 17 | 1 | 0 | 1 | 0 | 0 | 0 |
| 18 | 1 | 0 | 1 | 0 | 0 | 0 |
| 19 | 1 | 1 | 1 | 0 | 0 | 0 |
| 20 | 0 | 1 | 0 | 0 | 0 | 0 |
| 21 | 1 | 1 | 0 | 0 | 0 | 0 |
| 22 | 0 | 0 | 0 | 0 | 0 | 0 |
| 23 | 0.5 | 1 | 1 | 0 | 0 | 0 |
| 24 | 1 | 0 | 1 | 0 | 0 | 0 |
| 25 | 0 | 0 | 1 | 0 | 0 | 0 |
| 26 | 1 | 0 | 1 | 0 | 0 | 0 |
| 27 | 1 | 0.5 | 1 | 0 | 0 | 0 |

**Supplementary Table 6:** Coupling coefficients and phase coefficients normalized to π for the results showed in Fig. 6 in the supplementary document. 2x2 Pauli-y, Pauli-z, and Hadamard transformations.

| TriLatt | KA | KB | KC | TA/pi | TB/pi | TC/pi |
|---|---|---|---|---|---|---|
| 1 | 1 | 1 | 1 | 0 | 0 | 0 |
| 2 | 1 | 1 | 1 | 0 | 0 | -0.5 |
| 3 | 1 | 0 | 1 | 0 | 0 | 0 |
| 4 | 1 | 0 | 1 | 0 | 0 | 0 |
| 5 | 1 | 0 | 1 | 0 | 0 | 0 |
| 6 | 1 | 0 | 1 | 0 | 0 | 0 |
| 7 | 1 | 0 | 1 | 0 | -1 | -0.5 |
| 8 | 1 | 0 | 1 | 0.5 | -1 | 0 |
| 9 | 1 | 0 | 1 | 0 | 0 | 0.5 |
| 10 | 1 | 1 | 1 | 0 | 0 | 0 |
| 11 | 1 | 0 | 1 | 1 | 0 | -1.5 |
| 12 | 1 | 0.5 | 1 | 0 | 0 | 0 |
| 13 | 1 | 0 | 1 | 1.5 | 0 | 1 |
| 14 | 1 | 0 | 1 | 0 | 0 | 0 |
| 15 | 1 | 0 | 1 | 0 | 0 | 0 |
| 16 | 1 | 0 | 1 | 0.5 | 0 | 0 |
| 17 | 1 | 0 | 1 | 0 | 0 | 0.5 |
| 18 | 1 | 0 | 1 | 0 | 0 | 0 |
| 19 | 1 | 0 | 1 | 0 | 0 | 0 |
| 20 | 1 | 0 | 1 | 0 | 0 | 0 |
| 21 | 1 | 0 | 1 | 0 | 0 | 0 |
| 22 | 1 | 0 | 1 | 0 | 0 | 0 |
| 23 | 1 | 0 | 1 | 0 | 0 | 0 |
| 24 | 1 | 0 | 1 | 0 | 0 | 0 |
| 25 | 1 | 0 | 1 | 0 | -0.5 | 0 |
| 26 | 1 | 0 | 1 | 0 | 0.5 | 0 |
| 27 | 1 | 0 | 1 | 0 | 0 | 0 |

**Supplementary Table 7:** Coupling coefficients and phase coefficients normalized to π for the results showed in Fig. 7 in the supplementary document. DFT4 and CNOT transformations.

| TriLatt | KA | KB | KC | TA/pi | TB/pi | TC/pi |
|---|---|---|---|---|---|---|
| 1 | 1 | 0 | 1 | 0 | 0 | 0 |
| 2 | 1 | 0 | 1 | 0 | 0 | 1 |
| 3 | 1 | 0 | 1 | 0 | 0 | 1 |
| 4 | 1 | 0 | 1 | 0 | 0 | 1 |
| 5 | 1 | 0 | 1 | 0 | 0 | 1 |
| 6 | 1 | 0 | 1 | 0 | 0 | -1 |
| 7 | 1 | 1 | 1 | 0 | 0 | -1 |
| 8 | 1 | 0.5 | 1 | 0 | 0 | -1.5 |
| 9 | 1 | 0.5 | 1 | 0 | 0 | 0 |
| 10 | 1 | 0 | 1 | 0 | 0 | 0 |
| 11 | 1 | 0 | 1 | 0 | 0 | -1 |
| 12 | 1 | 0 | 1 | 0 | 0 | 1 |
| 13 | 1 | 0.6667 | 1 | 0 | 0 | -0.75 |
| 14 | 1 | 0 | 1 | 0 | 0 | 1 |
| 15 | 1 | 0 | 1 | 0 | 0 | 0 |
| 16 | 1 | 0 | 1 | 0 | 0 | 0 |
| 17 | 1 | 0.75 | 1 | 0 | 0 | -1.25 |
| 18 | 1 | 0.75 | 1 | 0 | 0 | -0.5 |
| 19 | 1 | 0 | 1 | 0 | 0 | 0 |
| 20 | 1 | 0 | 1 | 0 | 0 | 1 |
| 21 | 1 | 0 | 1 | 0 | 0 | 1 |
| 22 | 1 | 0.3334 | 1 | 0 | 0 | -0.5 |
| 23 | 1 | 0 | 1 | 0 | 0 | 1 |
| 24 | 1 | 0 | 1 | -2 | 0 | 0 |
| 25 | 1 | 0 | 1 | -2 | 0 | -1 |
| 26 | 1 | 0 | 1 | -1.75 | 0 | -2 |
| 27 | 1 | 0 | 1 | -2.25 | 0 | -1 |

**Supplementary Table 8:** Coupling coefficients and phase coefficients normalized to π for the results showed in Fig. 8 in the supplementary document. Hybrid 2x2 trasformation

| TriLatt | KA | KB | KC | TA/pi | TB/pi | TC/pi |
|---|---|---|---|---|---|---|
| 1 | 0 | 1 | 0 | 0 | 0 | 0 |
| 2 | 1 | 1 | 1 | 0 | 0 | 0 |
| 3 | 1 | 0.5 | 1 | 0 | 0 | -1.75 |
| 4 | 1 | 1 | 1 | 0 | 0 | 0 |
| 5 | 1 | 1 | 1 | 0 | 0 | 0 |
| 6 | 1 | 1 | 1 | 0 | 0 | 0 |
| 7 | 1 | 0 | 1 | -0.75 | -1.5 | 0 |
| 8 | 1 | 0 | 0.1716 | 0 | -0.5 | 0.25 |
| 9 | 1 | 1 | 1 | 0 | 0 | 0 |
| 10 | 1 | 1 | 1 | 0 | 0 | 0 |
| 11 | 1 | 1 | 1 | 0 | 0 | 0 |
| 12 | 1 | 0.5 | 1 | 0 | 0 | -0.5 |
| 13 | 1 | 1 | 1 | 0 | 0 | 0 |
| 14 | 1 | 1 | 1 | 0 | 0 | 0 |
| 15 | 1 | 1 | 1 | 0 | 0 | 0 |
| 16 | 1 | 0 | 1 | 0.125 | 0 | 0 |
| 17 | 1 | 0 | 1 | 0 | 0 | -0.125 |
| 18 | 1 | 1 | 1 | 0 | 0 | 0 |
| 19 | 1 | 1 | 1 | 0 | 0 | 0 |
| 20 | 1 | 1 | 1 | 0 | 0 | 0 |
| 21 | 1 | 0 | 1 | 0 | 0 | 0 |
| 22 | 1 | 1 | 1 | 0 | 0 | 0 |
| 23 | 1 | 1 | 1 | 0 | 0 | 0 |
| 24 | 1 | 1 | 1 | 0 | 0 | 0 |
| 25 | 1 | 0 | 1 | 0 | -0.5 | 0 |
| 26 | 1 | 0 | 1 | 0 | 0.5 | 0 |
| 27 | 1 | 1 | 1 | 0 | 0 | 0 |

**Supplementary Table 9:** Coupling coefficients and phase coefficients normalized to π for the results showed in Fig. 10 in the supplementary document. Multiple Rings with different cavity lengths.

| TriLatt | KA | KB | KC | TA/pi | TB/pi | TC/pi |
|---|---|---|---|---|---|---|
| 1 | 0 | 0.35 | 1 | 0 | 0 | 0 |
| 2 | 1 | 1 | 1 | 0 | 0 | 0 |
| 3 | 1 | 0 | 0 | 0 | 0 | 0 |
| 4 | 1 | 1 | 1 | 0 | 0 | 0 |
| 5 | 1 | 1 | 1 | 0 | 0 | 0 |
| 6 | 1 | 1 | 0 | 0 | 0 | 0 |
| 7 | 0 | 1 | 1 | 0 | 0 | 0 |
| 8 | 0 | 0 | 1 | 0 | 1.5 | 0 |
| 9 | 0 | 0.23 | 0 | 0 | 0 | 0 |
| 10 | 1 | 1 | 1 | 0 | 0 | 0 |
| 11 | 1 | 1 | 1 | 0 | 0 | 0 |
| 12 | 1 | 1 | 1 | 0 | 0 | 0 |
| 13 | 0 | 1 | 0 | 0 | 0 | 0 |
| 14 | 0 | 1 | 0 | 0 | 0 | 0 |
| 15 | 1 | 1 | 1 | 0 | 0 | 0 |
| 16 | 1 | 1 | 1 | 0 | 0 | 0 |
| 17 | 0 | 1 | 1 | 0 | 0 | 0 |
| 18 | 1 | 1 | 0 | 0 | 0 | 0 |
| 19 | 1 | 1 | 1 | 0 | 0 | 0 |
| 20 | 1 | 1 | 1 | 0 | 0 | 0 |
| 21 | 1 | 0 | 1 | 0 | 1 | 0 |
| 22 | 1 | 1 | 1 | 0 | 0 | 0 |
| 23 | 1 | 0.41 | 1 | 0 | 0 | 0 |
| 24 | 1 | 1 | 1 | 0 | 0 | 0 |
| 25 | 1 | 1 | 1 | 0 | 0 | 0 |
| 26 | 1 | 1 | 0 | 0 | 0 | 0 |
| 27 | 0 | 1 | 1 | 0 | 0 | 0 |

**Supplementary Table 10:** Coupling coefficients and phase coefficients normalized to π for the results showed in Fig. 11 in the supplementary document. Three-stage SCISSOR of two-CROWs

| TriLatt | KA | KB | KC | TA/pi | TB/pi | TC/pi |
|---|---|---|---|---|---|---|
| 1 | 0 | 0 | 0 | 0 | 0 | 0 |
| 2 | 0 | 0.07 | 1 | 0 | 0 | 0 |
| 3 | 0 | 0.07 | 0 | 0 | 0 | 0 |
| 4 | 1 | 0.07 | 0 | 0 | 0 | 0 |
| 5 | 0 | 0 | 0 | 0 | 0 | 0 |
| 6 | 1 | 0 | 0 | 0 | 0 | 0 |
| 7 | 0 | 0 | 0 | 0 | 0 | 0 |
| 8 | 0 | 0 | 0 | 0 | 0 | 0 |
| 9 | 0 | 0 | 1 | 0 | 0 | 0 |
| 10 | 0 | 0 | 0 | 0 | 0 | 0 |
| 11 | 0 | 0.07 | 1 | 0 | 0 | 0 |
| 12 | 0 | 0.07 | 0 | 0 | 0 | 0 |
| 13 | 1 | 0.07 | 0 | 0 | 0 | 0 |
| 14 | 0 | 0 | 0 | 0 | 0 | 0 |
| 15 | 1 | 0 | 0 | 0 | 0 | 0 |
| 16 | 0 | 0 | 0 | 0 | 0 | 0 |
| 17 | 0 | 0 | 0 | 0 | 0 | 0 |
| 18 | 0 | 0 | 1 | 0 | 0 | 0 |
| 19 | 0 | 0 | 0 | 0 | 0 | 0 |
| 20 | 0 | 0.07 | 1 | 0 | 0 | 0 |
| 21 | 0 | 0.07 | 0 | 0 | 0 | 0 |
| 22 | 1 | 0.07 | 0 | 0 | 0 | 0 |
| 23 | 0 | 0 | 0 | 0 | 0 | 0 |
| 24 | 1 | 0 | 0 | 0 | 0 | 0 |
| 25 | 0 | 0 | 0 | 0 | 0 | 0 |
| 26 | 0 | 0 | 0 | 0 | 0 | 0 |
| 27 | 0 | 0 | 1 | 0 | 0 | 0 |